\documentclass[english]{iopart}

\usepackage[T1]{fontenc}
\usepackage[latin1]{inputenc}
\usepackage{babel}
\usepackage{amsfonts}
\usepackage{amssymb}
\usepackage{graphics}
\usepackage{graphicx}
\usepackage[T1]{fontenc}
\usepackage[latin1]{inputenc}
\usepackage{epsf}

\def\is{{\bf I}-symmetries}
\def\be{\begin{eqnarray}}
\def\ee{\end{eqnarray}}
\newdimen\figwidth
\figwidth=\textwidth

\begin{document}

\title{ A classification of four-state spin edge Potts models}
\author{J.-Ch.~Anglès d'Auriac \dag, J.-M.~Maillard\ddag,
C.M.~Viallet\ddag\ } 
\address{\dag\ Centre de Recherches sur les
Tr\`es Basses Temp\'eratures, BP 166, 38042 Grenoble, France}
\address{\ddag\ LPTHE, Tour 16, 1er \'etage, Bo\^\i te 126, 4 Place
Jussieu, 75252 Paris Cedex 05, France} 
\ead{dauriac@grenoble.cnrs.fr,
maillard@lpthe.jussieu.fr, viallet@lpthe.jussieu.fr}

\begin{abstract}
We classify four-state spin models with interactions along the edges
according to their behavior under a specific group of symmetry
transformations. This analysis uses the measure of complexity of the
action of the symmetries, in the spirit of the study of discrete
dynamical systems on the space of parameters of the models, and aims
at uncovering solvable ones. We find that the action of these
symmetries has low complexity (polynomial growth, zero entropy). We
obtain natural parametrizations of various models, among which an
unexpected elliptic parametrization of the four-state chiral Potts
model, which we use to localize possible integrability conditions
associated with high genus curves.
\end{abstract}

\begin{indented}
\item[]{Keywords:  Integrability, spin-edge models, star-triangle
relations, inversion relations, birational transformations,
complexity, entropy.}
\end{indented}

\section{Introduction}

Our purpose is to describe the structure of the set of  spin
models with interactions along the edges, and propose a classification
of the matrix of Boltzmann weights of such models.

The main tool we use is the action of a group of transformations which
was exhibited in previous works concerning integrable
models~\cite{BeMaVi91c,BeMaVi91d,BeMaVi91e}, and which has its origin
in the so called inversion relations~\cite{Ba81}.  They are generated
by basic involutions, realized as birational transformations of the
parameters of the models (we will call this group of transformations
{\is}).

We analyze the behaviour of the realizations of {\is}, especially the
iterates of specific infinite order elements of the group.  The
guiding principle is that a chaotic behaviour of the realization of
{\is}, signaled by a measure of their complexity (entropy), is
preventing integrability of the corresponding statistical mechanical
model.

We concentrate on the four state models because of their richness, but
the method applies to all values of the number $q$ of states.

The paper is organized as follows: we describe the models we have in
mind and the action of {\is}. We then describe the classification of
the models according both to the dimension of their parameter space
and to their behavior under \is. We recover, in the course of this
classification, a number of known two-dimensional models which have
been studied in the literature, among which the standard scalar Potts
model~\cite{Wu82}, the Ashkin-Teller model~\cite{AsTe43}, the
Kashiwara-Miwa model~\cite{KaMi86}, and the four-state chiral Potts
model~\cite{HoKaNi83}.  One section is devoted to the analysis of
definite models.  Among other results, we give an
unexpected\footnote{Especially in view of the occurrence of higher
genus curves in the solutions of the star-triangle
relations~\cite{AuCoPeTaYa87,BaPeAu88}} elliptic parametrization of the
chiral Potts model. We give a constructive way to write explicitly the
algebraic locus where the group of {\is} degenerates into a finite
group, leaving room to solvability.  We finally comment on the
specificity of four-state models as compared to three and five or more
states.

\section{Spin edge models,  {\is},  and admissible patterns}
\label{models}

The models are defined on a $D$-dimensional regular lattice (square,
triangle, honeycomb, cubic, \dots), each vertex bearing a spin
$\sigma$ which can take $q$ values.  A model is defined by giving the
matrix of Boltzmann weights of local edge configurations: the state of
an edge is determined by the values of the spin at its ends. One
arranges the weights in a matrix $W$ whose entry $W_{i,j}$ is the
weight of the edge configuration with one end in state $i$ and the
other end in state $j$.  Notice that edges are oriented, and $W$ {\em
is not necessarily symmetric}. One may furthermore distinguish
different types of bonds, like, for example, vertical and horizontal
ones on a square lattice, and the most general non isotropic model is
then associated to a number of $q\times q$ matrices $W_\nu$, one for
each type of bond.

Since the entries of the matrices $W$ are Boltzmann weights, $W$ is
defined up to an overall multiplicative factor. The most general
matrix is then a collection of $q^2$ homogeneous entries, defined up
to a factor.

Simple transformations may be defined, which act on the matrices $W$,
forming the group of {\is}. This group is generated by simple
involutive generators, namely the matrix inverse $I$, and the element
by element inverse $J$, both taken up to a overall multiplicative
factor.  They may be written \be {I} : \quad W_{kl} \longrightarrow
A_{kl} \\ {J}: \quad W_{kl} \longrightarrow 1/W_{kl} \ee where $
A_{kl}$ is the cofactor of $ W_{kl}$ in $W$. To have a well defined
action of $J$, we assume that entries are non vanishing.

The two generators are non commuting involutions. They generate an
{\em infinite} group isomorphic to the dihedral group $Z_2 \ltimes Z$.
Its infinite part is made of the iterations of $\varphi= {I}\cdot {J}$
and its inverse $ \varphi^{-1}= {J}\cdot {I}$.

We will build families of models by successive reductions of the
number of parameters, obtained by imposing equality relations on the
entries of the Boltzmann matrix $W$.

To illustrate this, consider the chiral Potts model: its Boltzmann weight
 matrix  is a  cyclic matrix $W$, where instead of keeping all $q^2$ 
entries independent, one imposes the following  
equality conditions:
\be \label{cyclic}
W_{u,v} = W_{u+1, v+1}
\ee
getting for $q=4$ a pattern of the form:
\be \label{cyclic_matrix}
W = \pmatrix{  {\it w_0}&{\it w_1}&{\it w_2}&{\it w_3}\cr
	{\it w_3}&{\it w_0}&{\it w_1}&{\it w_2} \cr
	{\it w_2}&{\it w_3}&{\it w_0}&{\it w_1} \cr
	{\it w_1}&{\it w_2}&{\it w_3}&{\it w_0} \cr }
\ee
instead of the general $4\times 4 $ matrix. The number of parameters 
has gone down from $16$  to $4$ because we have imposed $12$  equality
relations.

The specificity of this type of relations is two-fold.  Any relation
of the form $W_{a,b} = W_{a',b'}$ for some pairs of indices $
\{a,b\}$, $\{a',b'\}$ defines a pattern of matrix which is
automatically left stable by the generator $J$ of the group of {\is}.

Some pattern defined in this way will also be left stable by the
matrix inversion $I$, {\em but not all of them}. We will say that a
pattern is {\em admissible} if its form is left stable by matrix
inversion~\cite{BeMaVi91c}.

Giving a pattern is equivalent to giving a partition of the $q^2$
entries of $W$. The number of possible patterns, that is to say the
number $P(q^2)$ of partitions of $q^2$ objects, is given
by~\cite{BeMaVi91c}
\begin{equation}
         {P}(q^2)=\sum_{s=1}^{q^2}\sum_{k=0}^{s-1} (-1)^k
                        {(s-k)^{(q^2-1)} \over k!\,(s-1-k)!}
\end{equation}
It grows extremely rapidly with the number $q$ of spin states. It is
$21147$ for $q=3$, $10480142147\simeq 10^{10}$ for $q=4$, and
$4638590332229998592\simeq 4.6 \cdot 10^{18} $ for $q=5$.

Our first step is to determine which of the $P(q^2)$ patterns are
admissible. For $q=4$ this was done by direct inspection of the
$P(16)\simeq 10^{10}$ patterns. The outcome is a list of $166$
patterns. We have not considered vanishing conditions on the entries,
and we restrict ourselves to invertible matrices~\footnote{Neither
have we considered equality up to sign between entries~\cite{Ba83}.},
so that both $I$ and $J$ are defined and invertible.

There are trivial redundancies in this list because any of the $q!$
permutations of the $q$ values of the spin, i.e. the same permutation
acting on simultaneously on the rows and columns of $W$ will take an
admissible pattern into an admissible pattern: it acts by similarity
on $W$, and commutes with {\is}.  This reduces the list to only $42$
different admissible patterns, given in   \ref{appa}.

We see that there are no admissible patterns with $9$, $11$, $12$,
$13$ , $14$, or $15$ free (homogeneous) parameters.

A matrix of the form (\ref{cyclic_matrix}) of the previous section
belongs to the similarity class of pattern ${\# 17}$, which contains
three elements (see later):
\begin{eqnarray*}
\hskip -2truecm
\pmatrix{
  {  w_0}&{  w_1}&{  w_2}&{  w_3}
\cr \noalign{\medskip}{  w_3}&{  w_0}&{  w_1}&{  w_2}
\cr \noalign{\medskip}{  w_2}&{  w_3}&{  w_0}&{  w_1}
\cr \noalign{\medskip}{  w_1}&{  w_2}&{  w_3}&{  w_0}
}, \qquad
\pmatrix{
  {  w_0}&{  w_1}&{  w_2}&{  w_3}
\cr \noalign{\medskip}{  w_2}&{  w_0}&{  w_3}&{  w_1}
\cr \noalign{\medskip}{  w_1}&{  w_3}&{  w_0}&{  w_2}
\cr \noalign{\medskip}{  w_3}&{  w_2}&{  w_1}&{  w_0}
}, \qquad
\pmatrix{
 {  w_0}&{  w_1}&{  w_2}&{  w_3}
\cr \noalign{\medskip}{  w_1}&{  w_0}&{  w_3}&{  w_2}
\cr \noalign{\medskip}{  w_3}&{  w_2}&{  w_0}&{  w_1}
\cr \noalign{\medskip}{  w_2}&{  w_3}&{  w_1}&{  w_0}
}
\end{eqnarray*}
These patterns have four homogeneous independent parameters $[w_0,
w_1, w_2, w_3]$.  Since they are admissible, there is an action
of the homogeneous matrix inversion $I$ on $[w_0, w_1, w_2, w_3]$,
which reads, when written for the first of the three representatives
(cyclic matrix):
\begin{eqnarray} \label{ipotts}
w_0 &\longrightarrow& w_0^3 - 2\; w_0 w_1 w_3 - w_0 w_2^2+ w_2 w_3^2 +
w_1 ^2 w_2 \nonumber \\ w_1 & \longrightarrow& -w_1^3 + 2\; w_0 w_1
w_2 + w_1 w_3^2 - w_3 w_0^2 - w_2 ^2 w_3 \\ w_2 & \longrightarrow&
w_2^3 - 2\; w_1 w_2 w_3 - w_2 w_0^2+ w_0 w_1^2 + w_3 ^2 w_0 \nonumber
\\ w_3 & \longrightarrow& -w_3^3 + 2\; w_0 w_2 w_3 + w_3 w_1^2 - w_1
w_2^2 - w_0 ^2 w_1 \nonumber
\end{eqnarray} \label{jpotts}
The action of $J$ reads, if we use the homogeneity of  the entries:
\begin{eqnarray*}
  [w_0, w_1, w_2, w_3]
 \longrightarrow
[\, { w_1}\,{ w_3}\,{ w_2},\quad { w_0}\,{ w_3}\,{ w_2},\quad { w_0}\,{
 w_1}\,{ w_3},\quad { w_1}\,{ w_2}\,{ w_0}]
\end{eqnarray*}

\section{Complexity analysis}
\label{complexity}

Since {\is} are essentially the iterates of $\varphi={I}\cdot {J} $, their
action looks like a discrete time dynamical system defined on the
parameter space of the model. The measure of complexity we use is
given by the rate of growth of the degrees of the iterates of
$\varphi$. Indeed once we have reduced the action of ${I}$ and ${J}$ to
the pattern, the action of $\varphi$ is a polynomial transformation on
the independent parameters. All expressions have a definite degree
$d$. The $n$-th iterate naively has degree $d^n$, but since we work
with homogeneous coordinates, we should factor out any common factor,
so that the degree may drop to a lower value $d_n$.

A measure of the complexity of the map is given by the
entropy~\cite{FaVi93,BeVi99} \be \epsilon = \lim_{n \rightarrow
\infty}\frac1n \log(d_n).  \ee

If $\epsilon =0$ the growth of $d_n$ is polynomial in $n$ and $d_n$ behaves
like 
\be
d_n=\alpha \;  n^\kappa \;  (1+ O(1/n))
\ee

and one may then define a secondary complexity index associated to the
map, the integer power $\kappa$.

The most compact way to encode the sequence of degrees $d_n$ is to
write its generating function:  \be g(s) = \sum_{n=0}^{n=\infty} d_n
s^n. \ee

This function was conjectured to always be {\em rational} with integer
coefficients, i.e. be of the form:
\begin{eqnarray}
g(s) = {P(s) \over{Q(s)}}
\end{eqnarray}
with $P$ and $Q$ some polynomials of degree $p$ and $q$ with integer
coefficients. It is sufficient in practice to calculate the $p+q$
first terms of the sequence to infer this generating
function\footnote{The catch is that the values of $p$ and $q$ are not
known in advance.}. Any further degree calculation serves as a
verification.

If the roots of the denominator of the generating function are all of
modulus one, then the entropy vanishes, and we may consider the
secondary index $\kappa$.

To illustrate the method, we may look at pattern $\# 17$ as in the previous
section. The degree of $\varphi$ is $9$. The first iterations of
$\varphi$ yield the sequence of degrees:

\be [1, 9, 33, 73, 129, 201, 289, 393, 513, 723, \dots] \ee 
The beginning of the sequence (up to $d_6$) may be fitted with the
generating function 
\be g(s) = {\frac { \left(1 + 3\,s \right) ^{2}}
{\left(1- s \right) ^{3}}} \ee which is compatible with the
above-mentioned conjecture. The subsequent terms are then be predicted
correctly, and $g_{17}$ is found.  We see that the growth of the
degree is polynomial ($\epsilon =0$) and the secondary index is
$\kappa=2$.

For each pattern the different representatives yield the same
transformations and consequently the same generating function.

If the entropy $\epsilon$ vanishes, the secondary index may be read
from the generating function $g$: if  $g(s)$ has a pole  of order
$\gamma$ at $s=1$, then $ \kappa +1 = \gamma $.

We have written the realizations of $\varphi = {I} \cdot {J} $ for the
admissible patterns, and calculated the generating function for all of
them.  {\em The result is that for all 42 admissible patterns the
growth of the degree is polynomial, i.e.  the entropy vanishes, and
the index $\kappa$ takes values $0, 1, 2$ and $3$} (see the following
sections, \ref{appa} and \ref{generating}).

The complexity can also be estimated using the following arithmetical
procedure: start with a matrix with random \emph{integer} entries, and
iterate $\varphi$, dividing all the coefficients by their greatest
common divisor.  The complexity can be estimated from the growth of a
typical entry.  In practice the number of digits of the entries grows
like $\lambda^n$, with $\epsilon = \log{\lambda}$.  Since the entries
of the matrix grow fast, one has to use special representation of the
integers allowing to manipulate arbitrary large values~\cite{gmp}. In
practice one performs the iteration with several initial random
integer matrices.  The average \( l(k) \) of the logarithm of the
entries is recorded as a function of the number of iterations \( k \).
The entropy \( \epsilon \) or the secondary index \( \kappa \) are
easily deduced from \( l(k) \).

\section{Structure of the set of admissible patterns}

There exists a partial order relation on the set of patterns, induced
by the partial order on the partitions of the entries. Indeed a
partition $\alpha = \{\alpha_1, \alpha_2, \dots, \alpha_\mu\}$ may be
finer than a partition $\beta = \{\beta_1, \beta_2, \dots,
\beta_\nu\}$, if all parts constituting $\alpha$ are subsets of the
constituents of $\beta$:
\begin{eqnarray}
 \alpha \prec \beta,\quad \mbox{ if }\qquad \forall k=1\dots \mu,\quad
\exists \; l\quad \mbox { s.t. }\quad \alpha_k \subset \beta_l 
\end{eqnarray} 
If $ \alpha \prec \beta$ will say that $\beta$ is a descendant of
$\alpha$. We obtain a descendant by adding further equality relations
on the entries, i.e. merging parts together.

As an illustration, pattern $\# 3$ i.e. the standard scalar Potts
model is a descendant of pattern $\# 17$ i.e. the chiral Potts model.
All patterns are descendant of pattern $\# 42$ which has the finest
partition.

The list of immediate descendants of the various patterns can be given
with the notation [ Pattern number, \{ list of direct descendants\}],
where $\{\}$ indicates an empty list.
\begin{eqnarray*}
\hskip -2truecm
& [4, \{ 2\}] , \quad 
[  5	,  \{   1 		\}] , \quad 
[  6	,  \{    		\}] , \quad 
[  7	,  \{   1,3 		\}] , \quad
 [  8	,  \{   3 		\}] , \quad 
[  9	,  \{   2 		\}] , \quad 
[  10 	,  \{   4,5 		\}] , \quad  &\\ \hskip -2truecm&
[  11	,  \{   4,6 		\}] , \quad
[  12	,  \{   4 		\}] , \quad 
[  13	,  \{   6 		\}] , \quad 
[  14	,  \{   5 		\}] , \quad 
[  15	,  \{   7 		\}] , \quad 
[  16	,  \{   5,7 		\}] , \quad   &\\ \hskip -2truecm&
[  17	,  \{   6,7 		\}], \quad
[  18	,  \{   9 		\}] , \quad 
[  19	,  \{   8 		\}] , \quad 
[  20	,  \{   5,6 		\}] , \quad 
[  21	,  \{   12 		\}] , \quad 
[  22	,  \{   8 		\}] , \quad     &\\ \hskip -2truecm&
[  23	,  \{   4,7 		\}] , \quad 
[  24	,  \{   9 		\}] , \quad   
[  25	,  \{   15 		\}], \quad
[  26	,  \{   16,17,20,25 		\}] , \quad   &\\ \hskip -2truecm& 
[  27	,  \{   13,14,20 		\}] , \quad   
[  28	,  \{   19,22 		\}] , \quad 
[  29	,  \{   12,15,23 		\}] , \quad 
[  30	,  \{   10,14,16,23 		\}] , \quad    &\\ \hskip -2truecm&
[  31	,  \{    11,13,17,23	\}] , \quad
[  32	,  \{   18,24 		\}] , \quad 
[  33	,  \{   10,11,20,21 		\}], \quad  &\\ \hskip -2truecm&
[  34	,  \{   8,9,23 		\}] , \quad  
[  35	,  \{   21,25,29 		\}] , \quad 
[  36	,  \{   23 		\}], \quad  &\\ \hskip -2truecm&
[  37	,  \{   26,27,30,31,33,35 	\}] \quad
[  38	,  \{   18,22,29,34,36 		\}] , \quad 
[  39	,  \{   24,30,34 		\}] , \quad   &\\ \hskip -2truecm&
[  40	,  \{   26, 30, 36 		\}] , \quad 
[  41	,  \{   19, 31, 34, 36 		\}], \quad
[  42	,  \{   28,32,37,38,39,40,41 	\}]&
\end{eqnarray*}

This yields the intricate graph of descent relations: 
\bigskip
\begin{center}
\includegraphics[width=\figwidth]{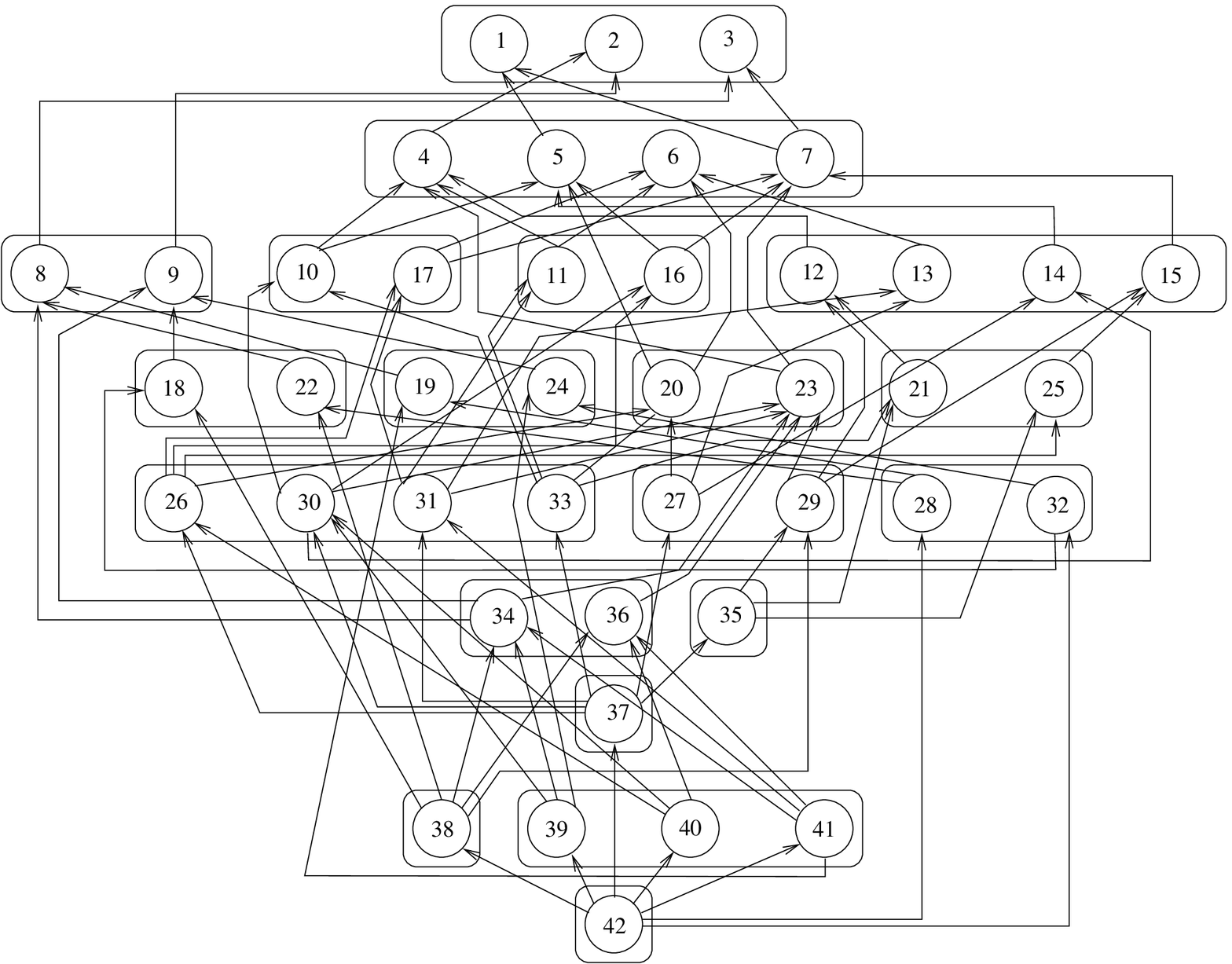}\\
Figure 1: Diagram of descent relations.
\end{center}

\section{Extended permutation classes}
\label{extended}

If two patterns $\Pi_1 $ and $\Pi_2$ verify $ \Pi_2 = L \cdot \Pi_1
\cdot R $, where both $L$ and $R$ are permutation matrices, the
realizations of $\varphi_1 = I_1 \cdot J_1 $ (resp.  $\varphi_2 = I_2
\cdot J_2 $) we get from $\Pi_1 $ (resp. $\Pi_2$) are again of the
same degree, even if $L$ and $R$ are not inverse of each other. We
even have $\varphi_1^2=\varphi_2^2$.  This explains why in the table
of generating functions, non similar patterns may  share the same
generating function. In the figures, we have used framing boxes to
indicate these equivalences, listed in the following table:
\begin{eqnarray*} \label{table_ext}
& \mbox{\# param.}\quad  &\qquad  \mbox{Classes}\qquad  \\
& 2	&   \Gamma_1 = \{ 1,2,3\}  \\
& 3 	&   \Gamma_2 = \{4,5,6,7\} \\
& 4	&   \Gamma_3= \{8,9\}, \quad \Gamma_4 = \{10, 17\}   \\
&	& \Gamma_5 = \{ 11, 16\}, \quad \Gamma_6 = \{12, 13, 14, 15\}  \\
& 5	&   \Gamma_7 = \{18,22 \}, \quad \Gamma_8 = \{19, 24\}  \\
&	& \Gamma_9 = \{20, 23\}, \quad \Gamma_{10} =\{21, 25\} 	\\
& 6	&   \Gamma_{11} = \{ 26 ,30, 31, 33 \}, \quad \Gamma_{12} = 
\{27, 29\}, \quad \Gamma_{13} = \{
28, 32 \}  \\
& 7	&  \Gamma_{14} = \{ 34, 36\}, \quad \Gamma_{15} =\{35\} \\
& 8	&  \Gamma_{16} =\{37\} \\
& 10	&   \Gamma_{17} = \{ 38 \}, \quad \Gamma_{18} = \{39, 40,41\}  \\
& 16	&  \Gamma_{19}=  \{42\}  \\
\end{eqnarray*}

Remark: $\varphi^2$ commutes with matrix transposition $t$ as well. A
a consequence, $t$ acts on the classes $\Gamma$. It actually acts
trivially on these classes.

\section{Gauge equivalence}
\label{gauge}

It is important to take into account the usual gauge symmetries of the
Boltzmann weights. Their action may be represented by a similarity
transformation: 
\begin{eqnarray}
\label{gauge_transf}
 W \longrightarrow W' = g^{-1} \cdot W \cdot g 
\quad \mbox{with} \quad
g = \pmatrix{ 	s_1 & 0 & 0 & 0 \cr
		0   & s_2& 0 & 0 \cr
		0   & 0 & s_3 & 0 \cr
		0 & 0 & 0 & s_4 }
\end{eqnarray}
They commute with the generator $I$ of {\is}, but not with $J$.
However, if two matrices verify (\ref{gauge_transf}), then their images
by $J$ are automatically gauge related by
\begin{eqnarray*}
g' = \pmatrix{ 	1/s_1 & 0 & 0 & 0 \cr
		0   & 1/s_2& 0 & 0 \cr
		0   & 0 & 1/s_3 & 0 \cr
		0 & 0 & 0 & 1/s_4 }.
\end{eqnarray*}
This means that $J$ has a well defined induced action on the gauge
equivalence classes, although it does not commute with the gauge
transformations.  Transformations as (\ref{gauge_transf})
yield the following equivalences:
\begin{eqnarray*}
 \Gamma_6=\{12, 13,14, 15\}  & \longrightarrow & \Gamma_2=\{4,5,6,7\}\\
 \Gamma_7 = \{18, 22\}  & \longrightarrow & \Gamma_3=\{8,9\} \\
\Gamma_{12} = \{27,29\} &  \longrightarrow & \Gamma_9=\{20,23\} \\
 \Gamma_{13}=\{28,32\} & \longrightarrow & \Gamma_8=\{19, 24\}\\
\end{eqnarray*}

Through these equivalences we have reductions by one unit of the
number of parameters. Moreover for all these reductions, there exists
an $I$-invariant which can be changed by the action of gauge
transformations, and the reduction amounts to fixing the value of this
invariant. Take for example pattern $\# 12$. One of its
representatives has Boltzmann matrix:
\begin{eqnarray*}
W_{12} = \pmatrix{ 
 {  w_0}&{  w_1}&{  w_2}&{  w_2}
\cr\noalign{\medskip}{  w_1}&{  w_0}&{  w_2}&{  w_2}
\cr\noalign{\medskip}{  w_3}&{  w_3}&{  w_1}&{  w_0}
\cr\noalign{\medskip}{  w_3}&{  w_3}&{  w_0}&{  w_1} }
\end{eqnarray*}
The quantity:
\begin{eqnarray*}
\Delta = {1\over{2}} \left( {{w_2}\over{w_3}} + {{w_3}\over{w_2}} \right)
\end{eqnarray*}
 is $I$-invariant. It can be brought to the value $\Delta=1$ by a
 gauge transformation of the form (\ref{gauge_transf}) with $s_1=s_2$
 , $s_3=s_4$, and $w_2\; s_3^2 = w_3 \;s_2^2$.  This makes $W_{12}' =
 g^{-1} \cdot W_{12} \cdot g$ symmetric and yields pattern  $\# 4$,
 with Boltzmann matrix:
\begin{eqnarray*}
W_{12}' = W_4 =  \pmatrix{ { w_0}&{ w_1}&{ w_2}&{ w_2}
\cr\noalign{\medskip}{ w_1}&{ w_0}&{ w_2}&{ w_2}
\cr\noalign{\medskip}{ w_2}&{ w_2}&{ w_1}&{ w_0}
\cr\noalign{\medskip}{ w_2}&{ w_2}&{ w_0}&{ w_1}}.
\end{eqnarray*}
We may picture the gauge equivalences by the following diagram:
\bigskip
\begin{center}
\includegraphics[width=\figwidth]{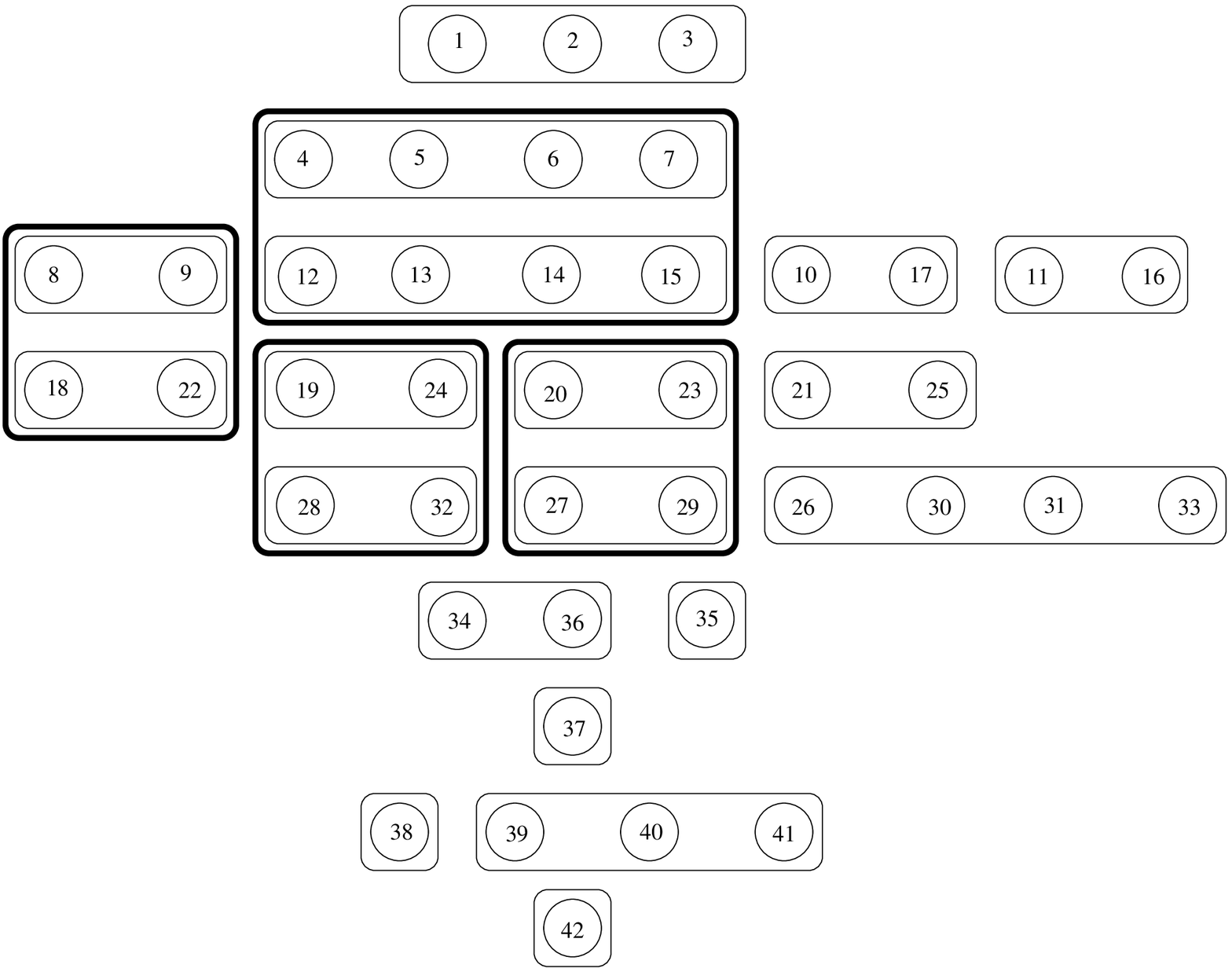}\\
Figure 2: Gauge reductions
\end{center}

The usual models which have been studied in the literature are present
in the top of this list (they have a low number of parameters). The
standard scalar Potts model is pattern $ \# 3$. The symmetric
Ashkin-Teller model is pattern $\# 7$.  The Ashkin-Teller (alias
Kashiwara-Miwa~\cite{KaMi86} with $N=4$) model is pattern $\# 16$.
The chiral Potts model is the pattern $\# 17$.

 They all share the property that {\is} have rational invariants,
 i.e. the orbits of the realization of {\is} are confined to algebraic
 varieties. Since we know how crucial is the role played by these
 varieties in the solvability of the models, it is natural to look for
 the structure of invariants of  the admissible patterns.

\section{Invariants of the action of {\is}}

Since all elements in an extended class yield the same birational
realizations, {\em up to a permutation of coordinates}, invariants may
associated to the classes $\Gamma$. 

The descendance of a given pattern being obtained by adding relations
among the entries of $W$, descendants inherit induced invariants
(rational or not). Some invariants will disappear or coalesce in the
process. What might also happen is that a transcendent invariant
degenerates to a rational one.

The value of $\kappa $ is related to the existence of conserved
algebraic varieties under the action of {\is}.  If there exist
invariant algebraic curves, $\kappa$ is necessarily $0,1$ or
$2$~\cite{Gi80,Be99}. If the curves are rational, $\kappa=0$ or
$1$. The values of $\kappa$ are shown on the following diagram.
\bigskip
\begin{center}
\includegraphics[width=\figwidth]{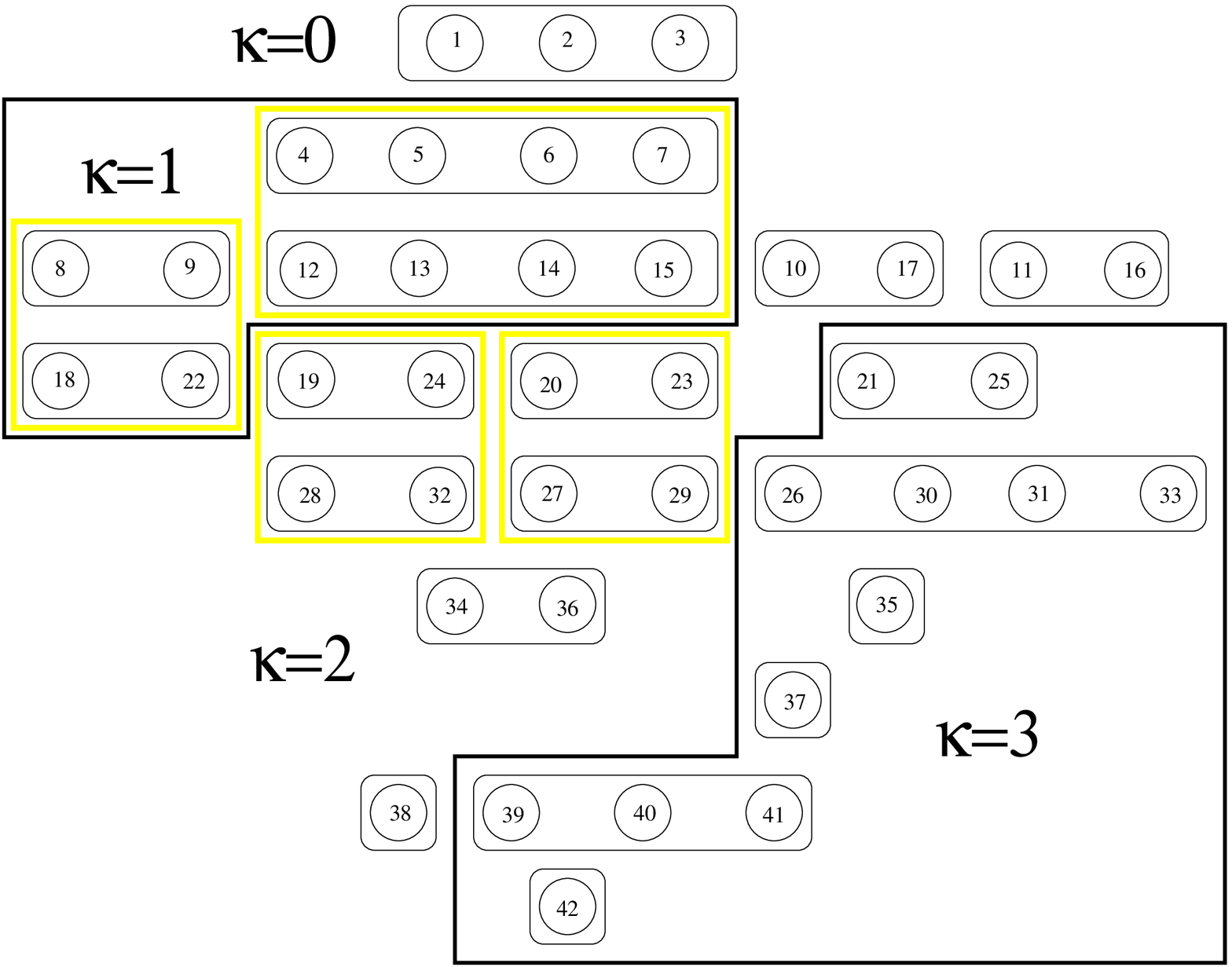}\\
Figure 3: Values of the index $\kappa$
\end{center}

All the values of \( \kappa \) are corroborated by the procedure
presented in paragraph 3. A number of iterations of the order of
thirty and an averaging over a number of the order of ten initial
matrices is enough to get a good precision on the value of \( \kappa
\).

There exists an unbounded algorithm to find rational
invariants~\cite{FaVi93}. There is however no simple way to discover
the non-rational ones, and it is an open problem to find all
$I$-invariants.  When the number of parameters is small, a graphical
approach may be extremely useful (see later with pattern $\# 25$).  In
the next section we will examine in detail the structure of typical
patterns.

\section{Analysis of specific patterns}

\subsection{Pattern $\#  3$: Standard scalar Potts model}
The Boltzmann matrix is: 
\begin{eqnarray*}
W_3 =  \pmatrix{ { w_0}&{ w_1}&{ w_1}&{ w_1} \cr\noalign{\medskip}{ w_1}&{
w_0}&{ w_1}&{ w_1} \cr\noalign{\medskip}{ w_1}&{ w_1}&{ w_0}&{ w_1}
\cr\noalign{\medskip}{ w_1}&{ w_1}&{ w_1}&{ w_0} }
\end{eqnarray*} 
In term of the parameter $w=w_1/w_0$, the action of $ \varphi= {I}\cdot
{J}$ reduces to a homographic transformation:
$
{I}: w \rightarrow -2 -w, \;
{J}:  w \rightarrow 1/w, \;
\varphi: w  \rightarrow -2 - 1/w
$.
In terms of the variable $ t=1/(1+w)$, $\varphi$ is the translation $
t \longrightarrow t-1$.

\subsection{Pattern $ \# 7$: Symmetric Ashkin-Teller model}
The Boltzmann matrix is: 
\begin{eqnarray*}
W_7 = \pmatrix{
 { w_0}&{ w_1}&{ w_2}&{ w_2}
\cr \noalign{\medskip}{ w_1}&{ w_0}&{ w_2}&{ w_2}
\cr \noalign{\medskip}{ w_2}&{ w_2}&{ w_0}&{ w_1}
\cr \noalign{\medskip}{ w_2}&{ w_2}&{ w_1}&{ w_0}
}
\end{eqnarray*}
and there is one algebraic invariant
\begin{eqnarray*}
\Delta_7 = {\frac {{ w_0}\,{ w_1}-{{ w_2}}^{2}}{{ w_2}\, \left( { w_1}-{ w_0} \right) }}
\end{eqnarray*}
The curves $\Delta_7=a$ have  a rational parametrization:
\begin{eqnarray*}
w_0=t , \quad w_1 = {{ 1-at}\over{t-a}}, \quad w_2=1
\end{eqnarray*}

\subsection{Pattern $\# 8$}
The  Boltzmann matrix is: 
\begin{eqnarray}
W_8 = \pmatrix{ { w_0}&{ w_1}&{ w_1}&{ w_1} \cr \noalign{\medskip}{
 w_1}&{ w_2}&{ w_3}&{ w_3} \cr \noalign{\medskip}{ w_1}&{ w_3}&{
 w_2}&{ w_3} \cr \noalign{\medskip}{ w_1}&{ w_3}&{ w_3}&{ w_2} }
\end{eqnarray}
The quantity
\begin{eqnarray}
\Delta_8^{(1)} = {\frac {{ w_2}\,{{ w_1}}^{2}+2\,{{
w_1}}^{2}{ w_3}-2\,{ w_0}\,{ w_3}\,{ w_2}-{ w_0}\,{{
w_3}}^{2}}{{ w_2}\,{{ w_1}}^{2}-{ w_0}\,{{ w_3}}^{2}}
}
\end{eqnarray}
is invariant by $J$ and changes sign under the action of $I$.  The
action of $\varphi$ exchanges the surfaces $\Sigma_+$
($\Delta_8^{(1)}=a$) and $\Sigma_-$ ($\Delta_8^{(1)}=-a$).  Since the
condition $\Delta_8^{(1)}=a$ may be solved rationally in $w_0$, we may
write the action of $\varphi^2$ on $\Sigma_+$, with coordinates $w_1,
w_2, w_3$.
In terms of the inhomogeneous variables $x = w_1/w_3$ and $y = w_3/
(w_2+w_3)$, it reads:
\begin{eqnarray}
 x & \longrightarrow & x'=\;  x \cdot  {\frac { \left( b+by-y \right)}
{ \left( 2+y \right) \left( by-2\,y+ b-1 \right) }}  \\ 
y & \longrightarrow & y' = \;  y +b
\end{eqnarray}
with $ b = 2a/(1+a)$, and the quantity
\begin{eqnarray}
\displaystyle
\Delta_8^{(2)}& =&  x \; \Phi(y) \quad \mbox{ with }\\
\Phi(y) &  = & 
{{\Gamma  \left( {\frac {y-2+b}{b}} \right) \Gamma  \left( {\frac {y+1}
{b}} \right)  }
/ 
 \Gamma  \left( {\frac {y-1+b}{b}} \right) 
/  \Gamma  \left( {\frac {y}{b}} \right) }
\end{eqnarray}
is invariant by the action of $\varphi^2$ on $\Sigma_+$.  This
example, where $\kappa=1$,  possesses a mixture of algebraic and
non-algebraic invariants, which can be evaluated exactly. The orbits
of $\varphi$ are confined to non-algebraic curves, and the n-th iterate may be written explicitly:
\begin{eqnarray}
y_n = y_0 + n \; b, \qquad x_n = x_0 \;{{\Phi(y_0)} \over{ \Phi(y_n)}}
\end{eqnarray}
The orbits accumulate to the point $ (x_\infty, y_\infty)=
(\Delta_8^{(2)}(x_0,y_0), \infty)$, which is a point on the line
$w_2+w_3=0$. 

 Pattern number $8$ leads to pattern number $3$ if $w_3=w_1$ and
$w_2=w_0$, in which case the above two invariants take definite
values: $\Delta_8^{(1)}=\infty$ and $\Delta_8^{(2)}=1$.

\subsection{Pattern $\# 16$: Ashkin-Teller  model}

This model coincides with the $N=4$ Kashiwara-Miwa
model~\cite{KaMi86}.  The Boltzmann matrix is:
\begin{eqnarray}
W_{16} = \pmatrix{
 { w_0}&{ w_1}&{ w_2}&{ w_3}
\cr \noalign{\medskip}{ w_1}&{ w_0}&{ w_3}&{ w_2}
\cr \noalign{\medskip}{ w_2}&{ w_3}&{ w_0}&{ w_1}
\cr \noalign{\medskip}{ w_3}&{ w_2}&{ w_1}&{ w_0}
}
\end{eqnarray}
There are two independent algebraic $I$-invariants
\begin{eqnarray}
\Delta_{16}^{(1)} ={\frac {{ w_1}\,{ w_3}-{ w_0}\,{ w_2}}{{
w_0}\,{ w_1}-{  w_3}\,{ w_2}}} , \qquad \Delta_{16}^{(2)} =
{\frac {{ w_1}\,{ w_3}-{ w_0}\,{ w_2}}{{ w_0}\,{ w_3}-{
 w_1}\,{ w_2}}}
\end{eqnarray}
Fixing the values of these two invariants to $a_1$ and $a_2$ determines 
an elliptic curve in the space of parameters. The equation of this curve is 
\begin{eqnarray}
 & a_{{1}}{x}^{2}-a_{{1}}{y}^{2}-a_{{2}}x-a_{{2}}a_{{1}}y+a_{{2}}{y}^{2}x
+a_{{2}}a_{{1}}y{x}^{2}=0 & 
\end{eqnarray}
\begin{eqnarray*} \hbox{ with } \quad
{\frac{w_1}{w_3}}=x, \quad { w_2\over{w_3}}=y, \quad
{w_0\over{w_3}} ={  {x+a_1 y}\over{y+a_1 x}} 
\end{eqnarray*}

\subsection{Pattern $\# 17$: Chiral Potts   model}
The chiral Potts model is equivalent to pattern $\# 17$. Its Boltzmann
matrix was already given above (equation (\ref{cyclic_matrix})). There
exist two algebraically independent invariants.
\begin{eqnarray}
\label{delta1}
\Delta_{Potts}^{(1)} &=& {\frac {{ w_3}\,{ w_2}-{ w_0}\,{ w_1}}{{
w_1}\,{ w_2}-{ w_0}\,{ w_3}}} \\ 
\label{delta2}
\Delta_{Potts}^{(2)} & = & {\frac {
\left( { w_0}\,{ w_2}-{{ w_1}}^{2} \right) \left( { w_0}\,{ w_2}-{{
w_3}}^{2} \right) }{ \left( { w_0}\,{ w_3}-{ w_1}\,{ w_2} \right)
^{2}}}
\end{eqnarray}
The condition $\Delta_{Potts}^{(1)}=a$ may be solved rationally in
$w_0$. Setting $\Delta_{Potts}^{(2)}= b$, together with $x=w_1/w_2$
and $y=w_3/w_2$, yields the equation of an elliptic curve to which the
orbit is confined:
\begin{eqnarray}
\hskip -2truecm
\label{potts_curve}
 \left( x-y \right) ^{2} \left( x+y \right) ^{2}b - \left( -y{x}^{2}a+
ax+{x}^{3}-y \right)  \left( -y+ax+{y}^{2}x-{y}^{3}a \right)=0 
\end{eqnarray}
We may illustrate the efficiency of the graphical method in this case,
since the set of the iterates of $\varphi$ is dense in a
curve\footnote{Contrary to what happens in the example of pattern $\#
8$ above.}.  Figure 4 shows a typical orbit in the three dimensional
space of parameters. The orbit is projected on a coordinate plane. The
aspect of the picture is very stable under changes of the starting
point of the iteration.
\bigskip
\begin{center}
\centerline {\epsffile[0 0 320 320]{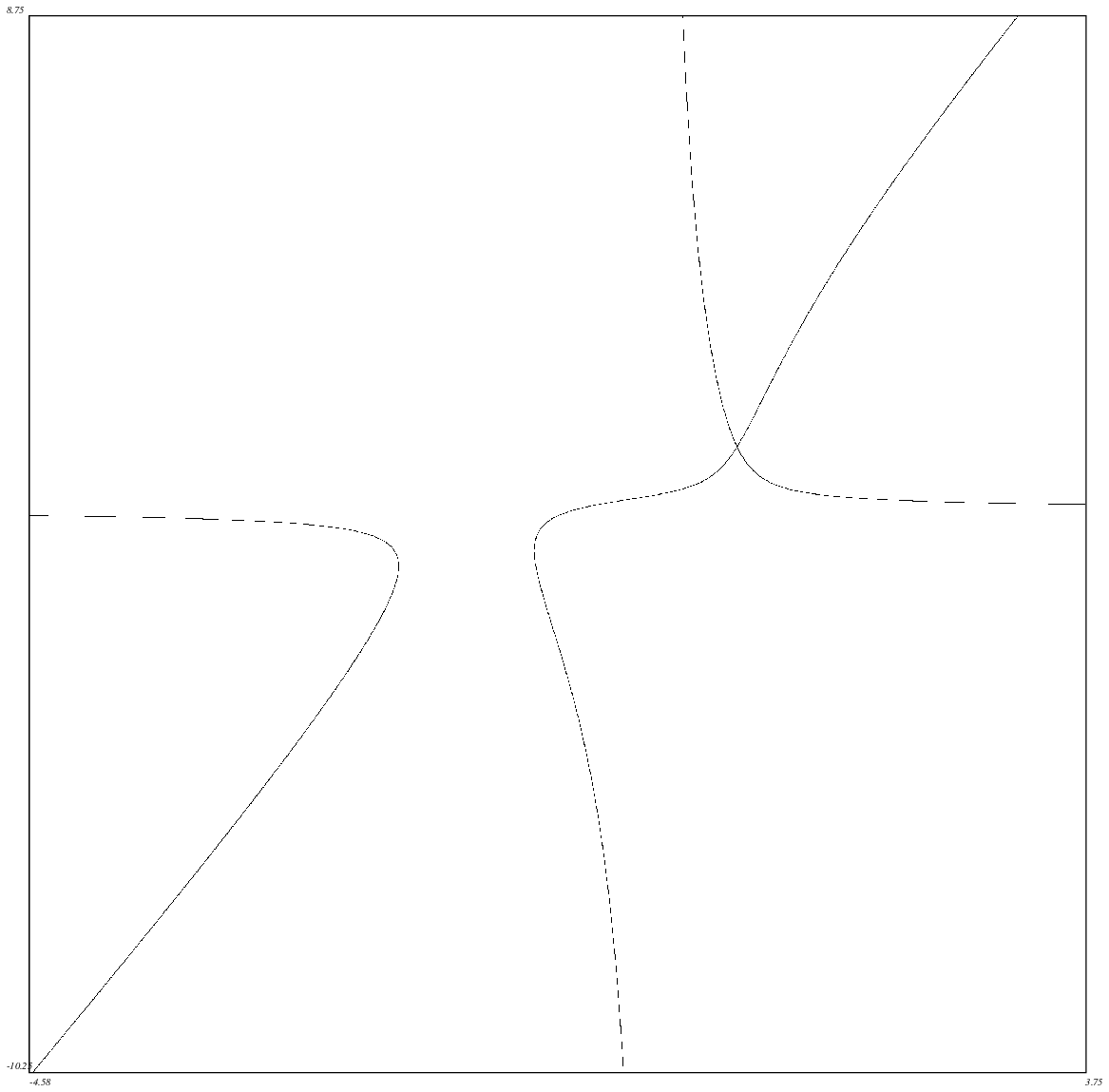} }
Figure 4: A typical orbit of $\varphi= I\cdot J$  for the chiral Potts model
\end{center}

The transformations $I$ and $J$ may be restricted to the quadrics
$a=cst$.  They may be written on $x$ and $y$, for fixed $a$.
\begin{eqnarray}
&{I}_a: &  x \longrightarrow  x' = -{\frac
{{y}^{3}-2\,{y}^{2}xa+2\,ya+{x}^{2}y{a}^{2}-{a}^{2}x-x}
{1 - a{y}^{2}+{a}^{2}yx-xy-{a}^{2}+a{x}^{2}}}
\\&& y  \longrightarrow  y'= {\frac
{{y}^{2}x-2\,{x}^{2}ay-y-y{a}^{2}+2\,ax+{x}^{3}{a}^{2}}
{1-a{y}^{2}+{a}^{2}yx-xy-{a}^{2}+a{x}^{2}}} \nonumber
\\& {J}_a: & x \longrightarrow  x'=1/x , \qquad  y  \longrightarrow  y'=1/y
\end{eqnarray}
The transformations ${I}_a$ and ${J}_a$ leave the curves
(\ref{potts_curve}) globally invariant {\em whatever $a$ and $b$ are}.

The modular invariant of the elliptic curve (\ref{potts_curve}) is
given by
\begin{eqnarray} \label{j}
j & = & {{1728}}\; {{g_2^3}\over{ g_2^3 - 27\; g_3^2}},\qquad \mbox{ with } \\
g_2 & = &1+ 256\,{b}^{4}-512\,a{b}^{3}-16\,{b}^{2}{a}^{4}+288\,
{b}^{2}{a}^{2}-16\,
{b}^{2}+16\,{a}^{5}b \nonumber \\\nonumber
&& -32\,{a}^{3}b  +16\,ab  -4\,{a}^{6}+{a}^{8}-4\,{a}^{2}+6\,{a}^{4} \\
g_3 &  =& \left(1 -32\,{b}^{2}+32\,ab-2\,{a}^{2}+{a}^{4}
 \right)  \left(1+ 16\,{b}^{2}-16\,ab-2\,{a}^{2}+{a}^{4} \right) \nonumber \\
&& \times \left(1 -8\,{b}^{2}+8\,ab-2\,{a}^{2}+{a}^{4} \right) /(3\sqrt{3})
\nonumber
\end{eqnarray}

The modular invariant (\ref{j})  can be written
\begin{eqnarray*}
j &=& 256 \; { (1-{\cal M}+{\cal M}^2)^3\over{ {\cal M}^2 \; ( 1-{\cal M})^2}}, \qquad
\end{eqnarray*}
where
\begin{eqnarray} \label{j2m}
{\cal M} = {(1-2a+a^2+4b)(1+2a+a^2-4b)\over{ (1-a)^2 (1+a)^2}}
\end{eqnarray}
is the square of the modulus of the elliptic functions parametrizing
the curve. Notice that ${\cal M}$ is ambiguously defined by condition
(\ref{j2m}), and may be replaced by any of the six values $\{ {\cal
M}, 1-{\cal M}, 1/{\cal M}, 1-1/{\cal M}, 1/(1-{\cal M}), {\cal
M}/({\cal M}-1) \}$.

There are  equivalent symmetric biquadratic forms of the curve
(\ref{potts_curve}):
\begin{eqnarray*}
\hskip -2truecm
 {p}^{2}{q}^{2}-2\, \left( {{\it J_x}}^{2}{{\it J_y}}^{2}+{{\it J_y}}^{2}{
{\it J_z}}^{2}+{{\it J_x}}^{2}{{\it J_z}}^{2} \right) p\, q+4\,{{\it J_x}}^{2
}{{\it J_y}}^{2}{{\it J_z}}^{2} \left( p+q \right)+ && \nonumber  \\
+ \left( {{\it J_x}}^{
2}{{\it J_y}}^{2}+{{\it J_y}}^{2}{{\it J_z}}^{2}+{{\it J_x}}^{2}{{\it J_z}}
^{2} \right) ^{2}-4\,{{\it J_x}}^{2}{{\it J_y}}^{2}{{\it J_z}}^{2}
 \left( {{\it J_x}}^{2}+{{\it J_y}}^{2}+{{\it J_z}}^{2} \right) = 0&&
\end{eqnarray*}
and
\begin{eqnarray*}
\hskip -2truecm
 \left( {\it J_x}-{\it J_y} \right)  \left( {p}^{2}{q}^{2}+1 \right) -
 \left( {\it J_x}+{\it J_y} \right)  \left( {p}^{2}+{q}^{2} \right) +4\,
{\it J_z}\,p\,q =0
\end{eqnarray*}
with
\begin{eqnarray*}
\hskip -2truecm
{\it J_x} = 2\, a, \qquad {\it  J_y} = 4\, b-2\, a, 
\qquad {\it  J_z}=a^2+1, \quad\mbox{
and }\quad {\cal M} = {{\it J_z}^2 - {\it J_y}^2 \over
{  {\it J_z}^2 -{ \it  J_x}^2}}
\end{eqnarray*}
Changing ${\cal M}$ to $1-{\cal M}$ or $1/{\cal M}$ amounts to
permuting ${\it J_x}$, ${\it J_y}$, and ${ \it  J_z}$.

Generic orbits of $\varphi= {I}\cdot {J}$ are infinite, but we know
how important are the degenerate cases where these orbits are finite:
for two-dimensional lattice models, this is where non trivial
star-triangle integrability takes
place~\cite{AuCoPeTaYa87,HaMa88}. For given values of $a$ and $b$, the
action of $\varphi$ is a shift on the curve (\ref{potts_curve}).

It is possible to write down the Weierstrass form of the curve
\begin{equation}
\label{weiercurve}
X^3 - \alpha \; X - \beta + Y^2 = 0
\end{equation}
with\footnote{The normalizations could be changed, and the powers of
$1/(b-a)$ could be absorbed in the definition of $X$ and $Y$.}
\begin{eqnarray}
\alpha = {1\over{48}} { g_2 \over{ (b-a)^8}}, \qquad
\beta = { \sqrt{3} \over { 288}} { g_3 \over{ ( b-a)^{12}}}
\end{eqnarray}
 and give an explicit coordinate transformation from the original
variables $(x,y)$ to $[X,Y]$.  We obtained the transformation through
Maple's implementation of the van Hoeij algorithm~\cite{Ho95} , after
setting $ b=a+1/u^2$.

The action of $\varphi$, when written on (\ref{weiercurve}) is the
addition\footnote{To add two points on the cubic (\ref{weiercurve}),
 draw the straight line through the two points. Compute the third
point of intersection of the line with the curve, and take its
symmetric under $Y\rightarrow -Y$.}  of the point $\Pi_+$ with
coordinates $[X_{\pi_+}, Y_{\pi_+}]$:
\begin{eqnarray*}
X_{\pi_+}& =&-1/12\;{\frac
{1+ 16\,{b}^{2}+8\,ab-2\,{a}^{2}+12\,b+12\,{a}^{2}b+{a}^ {4}}{ \left(
b-a \right) ^{4}}}\\ Y_{\pi_+}& =&1/2\;{\frac {b \left(1+ a \right) ^{2
} \left(1+ 4\,b-2\,a+{a}^{2} \right) }{ \left( b-a \right) ^{6}}}
\end{eqnarray*}
Although the change of coordinates depends on $u$, the $[X,Y]$
coordinates of the point $\Pi_+$ depend only on $u^2$ and may be
written rationally in terms of $a$ and $b$.

The inverse map $\varphi^{-1}$ is the addition of the opposite point
$\Pi_- = [X_{\pi_+},-Y_{\pi_+}]$.  Notice that both $\Pi_+$ and
$\Pi_-$ are obtained from the double point $x=-1/au, y=\infty$ of
(\ref{potts_curve}) in the desingularisation process which leads to
the Weierstrass form (\ref{weiercurve}).
 
The sum of $\Pi_\pm$ with itself, denoted $ 2\cdot\Pi_\pm$, is given
by
\begin{eqnarray}
\label{wpi2}
\hskip -1 truecm
[-1/12\,{\frac {1+{a}^{4}+10\,{a}^{2}-16\,ab+16\,{b}^{2}}{ \left( b-a
 \right) ^{4}}},\pm 1/2\,{\frac { \left(1+ {a}^{2} \right)  \left( 2\,b-
a \right) a}{ \left( b-a \right) ^{6}}}]
\end{eqnarray}
and the multiples $n\cdot\Pi_\pm$ are then easily obtained by
 recurrence.  Their $[X,Y]$ coordinates depend only on $u^2$ and may
 be re-expressed rationally in terms of $a$ and $b$.

The point $2\cdot \Pi_+$ corresponds to the point
\begin{eqnarray*}
(x,y) = ({\frac { \left( 1+{a}^{2} \right)
au}{1 - {u}^{2}{a}^{3}-{a}^{2}}},{ \frac { \left( 2+a{u}^{2} \right)
{a}^{2}u}{1-{u}^{2}{a}^{3}-{a}^{2}}} )
\end{eqnarray*}
of the original curve (\ref{potts_curve}).

The condition $\varphi^{(m+n)} = id$ is just:
\begin{eqnarray*}
X( n\cdot \Pi_+ )= X(m\cdot \Pi_-) \qquad \mbox{  and }\quad  
Y( n\cdot \Pi_+ )= -Y(m\cdot \Pi_-)
\end{eqnarray*}

As an example, the condition that $\varphi$ is of order $3$ is:
\begin{eqnarray}
v_3 = (1+a)^2 \; b - a^2 = 0
\end{eqnarray}
which is an algebraic condition on the entries of the Boltzmann matrix,
 and can be checked directly.

The condition that $\varphi$ is of order 4 is the vanishing of
$Y(2\cdot \Pi_+)$ given in (\ref{wpi2}) i.e.
\begin{eqnarray*}
v_4 =\left( 1+{a}^{2} \right)  \left( 2\,b- a \right) a
\end{eqnarray*}
The condition $2\; b - a = 0$ is precisely the condition appearing
in~\cite{AuCoPeTaYa87,HaMa88}, where solutions of the star-triangle
relations related to higher genus curves were exhibited.  The
condition $v_3=0$ is not known to yield similar integrability
conditions and was not found to yield any transition in the statistics
for the Random Matrix Theory approach (see~\cite{AnMaVi02}).

\subsection{Pattern $\# 25$}
This example is a good illustration of the mixed use of algorithmic
research of algebraic invariants, graphical method, and the knowledge
of the entropy and index $\kappa$.  The Boltzmann matrix is:
\begin{eqnarray}
W_{25} = \pmatrix{
 { w_0}&{ w_1}&{ w_2}&{ w_3}
\cr \noalign{\medskip}{ w_4}&{ w_0}&{ w_4}&{ w_2}
\cr \noalign{\medskip}{ w_2}&{ w_3}&{ w_0}&{ w_1}
\cr \noalign{\medskip}{ w_4}&{ w_2}&{ w_4}&{ w_0}
}
\end{eqnarray}
The number of inhomogeneous parameters is four. We found two
independent algebraic invariants.
\begin{eqnarray}
\Delta_{25}^{(1)}& =& {\frac {{ w_0}\,{ w_1}-{ w_3}\,{ w_2}}{{ w_0}\,{ w_3}-{ w_1}\,{ w_2}}}
\\
\Delta_{25}^{(2)}& =& {\frac { \left( { w_4}\,{ w_3}-{{ w_0}}^{2} \right)  \left( { w_1}\,{ w_4}-{{ w_2}}^{2} \right) }{ \left( { w_4}\,{ w_3}-{{ w_2}}^{2} \right)  \left( { w_1}\,{ w_4}-{{ w_0}}^{2} \right) }} 
\end{eqnarray}
Drawing orbits of the iterates of $\varphi$ leads, with a proper
choice of starting point, to pictures like Figure 5. Contrary to what
happens with pattern $\# 17$ (Figure 4), the shape of the orbit is
rather unstable under changes of the starting point.
\bigskip
\begin{center}
\centerline {\epsffile[0 0 320 320]{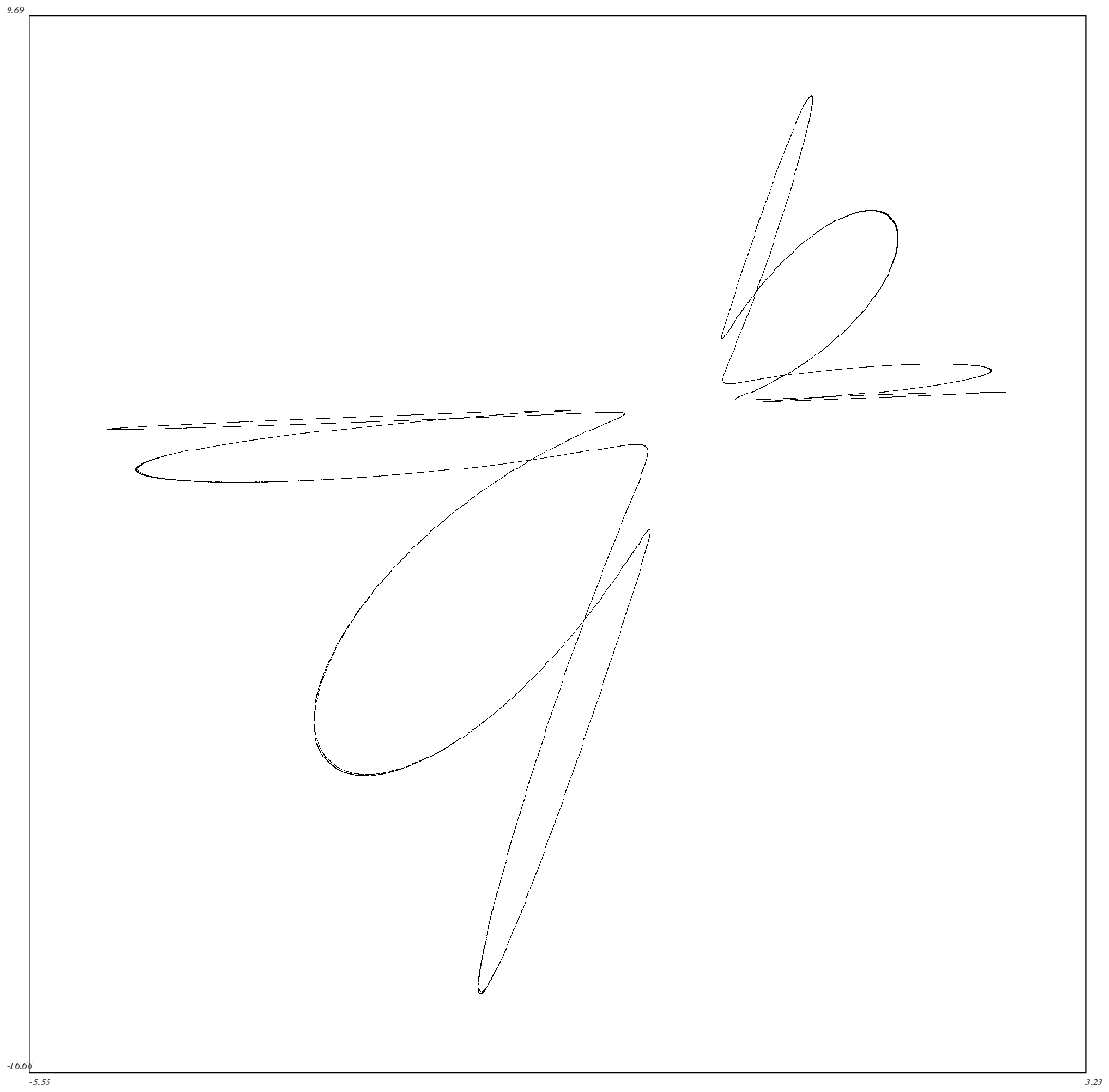} }
Figure 5: A typical orbit of $I$-symmetries for pattern $\# 25$
\end{center}
Such a picture indicates that there exists another independent
invariant. Would that invariant be algebraic, the stable curves would
be elliptic, and we know that the index $\kappa$ would then be less
than $3$. Since $\kappa=3$ we know this additional invariant has to be
non-algebraic.

Equation of the surface $\Sigma_1: \Delta_{25}^{(1)}=a$ may be solved
for $w_1$. We get an expression of $I$ and $J$ on $\Sigma_1$.
Changing variables to 
\begin{eqnarray*}
[x,y,z,t] = [ w_0^2, w_0 w_2, w_0 w_3 , w_3 w_4]
\end{eqnarray*}
 we obtain:
\begin{eqnarray} \label{i_a}
\hskip -2truecm
{I}_a([x,y,z,t])  = 
&& [\; x \left( x+ay-at-t \right) ^{2} \left( x+y \right)  \left( x+ay
 \right) , \nonumber \\ 
&& \left( x+y \right)  \left( -yx+xta+xt-{y}^{2}a \right) 
 \left( x+ay \right)  \left( x+ay-at-t \right) , \nonumber \\
 &&-z \; \left( x+y \right) 
 \left( {x}^{2}+xt{a}^{2}-xt-{y}^{2}{a}^{2} \right)  \left( x+ay-at-t
 \right) , \nonumber\\
&& \left( x+ay \right)  \left( x-y \right) t \left( {x}^{2}+xt
{a}^{2}-xt-{y}^{2}{a}^{2} \right) ]
\end{eqnarray}
with remaining algebraic invariant
\begin{eqnarray}
\Delta_{25}^a = {\frac {x \left( t-y \right) \left(yt -yx-{y}^{2}a+axt
\right) }{ \left( x-y \right) ^{2}t \left( x+y \right) }}
\end{eqnarray}
Notice that in equation (\ref{i_a}), the third coordinate $z$ is just
multiplied by a factor depending on the other coordinates, and does
not appear anywhere else in the induced transformation ${I}_a$ nor in
the invariant $\Delta_{25}^a$. The curves $\Delta_{25}^a =b$ are
elliptic curves in the variables $(x,y,t)$, extending to cylinders in
$[x,y,z,t]$. The curve appearing on Figure 5 is drawn on such a
cylinder.

We may examine a particular value of $b$ where the curve
$\Delta_{25}^a =b$ degenerates to a rational curve, but there exists a
transcendent invariant for ${I}_a$. The rational parametrization of the
cylinder is
\begin{eqnarray}
[x,y,z,t] =  [{\frac {{s}^{2}t \left( 1+sa \right) }{s+a}},
{\frac {st \left( 1+sa \right) }{s+a}},z,t]
\end{eqnarray}
The induced map $({I}\cdot {J})^2$ may be written 
\begin{eqnarray}
&& S \longrightarrow q^2 \; S, \qquad z \longrightarrow z \; 
{T(S)\over{T(q\,S)}} \\
&&   \hbox{with} \qquad
T(S) = {(1-S)(1-q^3\; S)\over{ ( 1-q^2\; S) (1-q^5\; S)}} 
\qquad\mbox{and} \nonumber \\
&& S={s-q^{-1}\over{s-q}}, \qquad 
q = -1/2\,{\frac {a+1+\sqrt {(1-a) \,( 1+3 a)}}{a}}. \nonumber
\end{eqnarray}
If we introduce the infinite product
\begin{eqnarray} 
\Pi(S) =  \prod_{k=0}^{k=\infty} { T(q^{4k} \; S) \over { T(q^{4k+1}
\; S)}}
\end{eqnarray}
then the $2n$th iterate of $(i.j)$ reads 
\begin{eqnarray}
 S_{2n} = q^{4n} \; S_0,
\qquad z_{2n} = z_0 \cdot { \Pi(S_0)\over{ \Pi(S_{2n})}} 
\end{eqnarray}
 i.e.  the transcendent quantity $\Delta = z \cdot \Pi(S)$ is
invariant by $({I} \cdot {J})^2$. The convergence of these Eulerian
products is ensured when $ a \in ( -1/3, 1)$. Such values of $a$ would
produce orbits with accumulation points. If $a$ is outside this
interval, then $q$ has unit modulus and the orbits are similar to the
one shown in Figure 5. The quantity $\Pi(S)$ is replaced by the
standard analytic prolongation.

\section{Other values of $q$}

We have investigated the structure of the set of admissible patterns
for other values of $q$. 

The value $q=4$ is an important threshold. It is known to be one for
the standard scalar Potts model in statistical mechanics~\cite{Wu82},
and for the chromatic polynomials and coloring problems. What we show
is that it is {\em the maximal value ensuring the vanishing of
entropy, whatever admissible pattern we choose}.

In the case of a cyclic matrix of size $q$ (chiral Potts model), with
 entries $[w_0,\dots, w_{q-1}] $ there is a simple relation between
 matrix inverse and element by element inverse. Define the linear
 transformation $C$ (discrete Fourier transform, or Kramers-Wannier
 duality)
\begin{eqnarray*}
C: \qquad w_k \longrightarrow \sum_{r=0}^{q-1} \omega^{\;k \;r} w_r, 
\end{eqnarray*} 
where $\omega$ is the $q$-th root of unity. We then have:
\begin{eqnarray*}
{I} =  C^{-1} \cdot {J} \cdot C.
\end{eqnarray*}
 It is possible to analyze the singularity structure of the
transformations $I$ and $J$, and prove that the generating function for
the degree of the iterates of $\varphi$ for the chiral Potts model
with $q$ states is~\cite{BeVi99}
\begin{eqnarray} \label{genepotts}
f_q(s) = {\frac { \left(1+ s\;(q-1) \right) ^{2}}{ \left( 1-s \right)
 \left(1- s\,(q^2  - 4\, q + 2)+ {s}^ {2} \right) }}.
\end{eqnarray}
As soon as $q \geq 5$, the entropy becomes strictly positive for this
family of admissible patterns.

For all sizes $q$, there exists a pattern depending on $q(q+1)/2$
homogeneous parameters: the symmetric matrix.

An open question is to decide whether or not there exist patterns
depending on $r$ parameters, with $q(q+1)/2 < r < q^2$. For $q=3$ and
$q=4$ there is a no man's land between $q(q+1)/2$ parameters
(symmetric matrix) and $q^2$ parameters (generic $q\times q$ matrix).

Another open question is to know if there is anything between cyclic
and symmetric matrix when $q$ is prime.

The value $q=2$ is trivial and yields a finite group of symmetries.
When $q=3$, the same analysis can be performed completely (see below).
When $q=5$, the number of patterns is too large to be thoroughly
examined, and we have restricted ourselves to patterns with only three
homogeneous parameters.

Of course there exist admissible patterns with larger values of $q$
still yielding vanishing entropy. They may have for example a block
structure, the blocks being themselves admissible patterns of smaller
sizes~\cite{BeMaVi91a,BoMaRo94,MeAnMaRo94,MeAn95}.

\subsection{Three state models}

The analysis can be made along the same lines as for the four-state
models: enumeration, calculation of the entropy, calculations of
invariants, etc. It is  summarized  here.

The number of partitions of nine objects is $21147$ and the number of
admissible patterns is only $9$, when equivalences are taken into
account.  The list of patterns is given in \ref{three}. For patterns
$\# 1,\# 2,\# 3,\# 4$ the group of {\is} is finite. Patterns $\# 3$
and $\# 4$ are related by an extended permutation similarity as
defined in section (\ref{extended}).  Patterns $\# 5$ and $\# 6$ yield
the {\em same} generating function for the degrees of the iterates of
$\varphi$
\begin{eqnarray*}
g_5 = g_6 = {\frac { \left( 1+ 2\,s \right) \left(1+ s \right) ^{2}}{
\left(1+s+ {s}^{ 2} \right) \left( 1-s \right) ^{2}}},
\end{eqnarray*}
 and the trajectories of $\varphi$ are confined to algebraic curves.
Patterns $\# 7,\# 8,\# 9$ yield the generating function
\begin{eqnarray} \label{taxi}
g_7 = g_8 = g_9 = {\frac
{1+6\,s+9\,{s}^{2}+2\,{s}^{3}+6\,{s}^{6}+3\,{s}^{7}-6\,{s}^{8}-3\,{s}^{9}}{
\left(1+s+ {s}^{2} \right) \left( 1-s \right) ^{2}}}.
\end{eqnarray}
Patterns $\# 7$ and $\# 8$ (pattern $\# 8$ is the $3\times 3 $ symmetric
matrix) are also related by an extended permutation similarity.  A
list of algebraic invariants is given for the various patterns in
  \ref{inva_three}.

For pattern $\# 8$ (symmetric $3\times 3$ matrices), the Boltzmann matrix is
\begin{eqnarray}
W_8 = \pmatrix{
 { w_0}&{ w_1}&{ w_2} \cr
\noalign{\medskip}{ w_1}&{ w_3}&{ w_4}\cr
\noalign{\medskip}{
 w_2}&{ w_4}&{ w_5} }
\end{eqnarray}
 If we normalize the entries with the condition $w_0=1$, the action of
$\varphi^3$ reduces to the following homothetic transformation:

\begin{eqnarray*}
[1, w_1,w_2,w_3,w_4,w_5] \longrightarrow [ 1, \Delta_1 \; w_1,
\Delta_2 \; w_2, \Delta_3 \; w_3, \Delta_4 \; w_4, \Delta_5 \; w_5]
\end{eqnarray*}
with 
\begin{eqnarray*}
\Delta_1 = {\frac { \left( { w_5}-{{ w_2}}^{2} \right) { w_3}\,{
 w_4}\, \left( -{ w_1}\,{ w_2}+{ w_4} \right) }{ \left( -{ w_3}\,{
 w_5}+{{ w_4}}^{2} \right) { w_2}\, \left( -{ w_2}\,{ w_3}+{ w_1}\,{
 w_4} \right) }} \\ \Delta_2 = {\frac { \left( { w_3}-{{ w_1}}^{2}
 \right) { w_5}\,{ w_4}\, \left( -{ w_1}\,{ w_2}+{ w_4} \right) }{
 \left( -{ w_3}\,{ w_5}+{{ w_4}}^{2} \right) \left( -{ w_1}\,{ w_5}+{
 w_4}\,{ w_2} \right) { w_1}}} \\ \Delta_3 = \Delta_1^2, \qquad
 \Delta_4 = \Delta_1 \; \Delta_2, \qquad \Delta_5 = \Delta_2^2
\end{eqnarray*}
where $\Delta_1$ and $\Delta_2$ are left invariant by the action of
$\varphi$.

The appearance of third root of unity in \ref{taxi} has to do with the
specificity of the third power of $\varphi$.

There is a parametrization of the general three-state model (pattern
$\# 9$) which simplifies the action of {\is}. The matrix of Boltzmann
weights reads
\begin{eqnarray*}
W =
\pmatrix{
w_0 & w_1 & w_2 \cr
w_3 & w_4 & w_5 \cr
w_6 & w_7 & w_8 \cr
}
\end{eqnarray*}
If one introduces the quantities
\begin{eqnarray*}
&& \alpha= w_0^2 ,\qquad \beta=w_4^2,\qquad \gamma=w_8^2 \\ && \delta
= w_1\; w_3,\qquad \epsilon = w_2\; w_6,\qquad \zeta= w_5\; w_7 \\ &&
\lambda = {w_1 \over{w_3}},\qquad \mu = {w_6\over{w_2}},\qquad \nu =
{w_5\over{w_7}}
\end{eqnarray*} 
The array $[ \alpha, \beta, \gamma, \delta, \epsilon, \zeta]$
transforms under  $\varphi^3$ by multiplication of its entries by factors
$[k_\alpha, k_\beta, k_\gamma, k_\delta, k_\epsilon, k_\zeta ]$. One
can check that all the ratios $k_\alpha/ k_\beta, \dots$ are
invariants of $\varphi^3$.  In a similar way, the three quantities
$[\lambda, \mu, \nu]$ transform under $\varphi^6$ by multiplication by
factors $[k_\lambda, k_\mu, k_\nu]$  which are all invariants of
$\varphi^6$.  

This yields a system of algebraic invariants of rank four of
$\varphi^3$, for example:
\begin{eqnarray*}
[{\frac {{ w_1}\,{ w_5}\,{ w_6}}{{ w_2}\,{ w_3}\,{ w_7}}}, \quad
{\frac {{ w_3}\,{ w_7}\,{ w_1}\,{ w_5}}{{{ w_4}}^{2}{ w_6} \,{ w_2}}},
\quad {\frac {{{ w_0}}^{2}{ w_5}\,{ w_7}}{{ w_1}\,{ w_2}\,{ w_3}\,{
w_6}}}, \quad {\frac {{{ w_8}}^{2}{ w_3}\,{ w_1}}{{ w_2}\,{ w_6}\,{
w_5}\,{ w_7}}}]
\end{eqnarray*}

\subsection{Five state models and beyond}

The number of partitions of 25 objects is over $4.10^{18}$. To
enumerate the configurations is beyond reach.  We have restricted our
analysis in the following way: if one looks for  partitions with only
three parts, that is to say five-state models with only three possible
colors (three homogeneous parameters), then there are ``only''
$141197991025 \simeq 141.10^{9}$ patterns. We have performed their
full enumeration, and found the subset of admissible ones.  The
interesting result is that there exits a unique admissible pattern, up
to permutations of rows and columns: the symmetric Potts model, with
symmetric cyclic $5\times 5 $ matrix of Boltzmann weights
\begin{eqnarray*}
W = \pmatrix {
w_0 & w_1 & w_2 & w_2 & w_1 \cr
w_1 & w_0 & w_1 & w_2 & w_2 \cr
w_2 & w_1 & w_0 & w_1 & w_2 \cr
w_2 & w_2 & w_1 & w_0 & w_1 \cr
w_1 & w_2 & w_2 & w_1 & w_0 \cr
}.
\end{eqnarray*}
There are actually $72$ representatives, all related by independent
permutations of rows and columns, as in section~(\ref{extended}).

The trajectories of the iterations of $\varphi$ are confined to
algebraic curves. The entropy vanishes and the secondary index
$\kappa$ has value $2$~\cite{Gi80,Be99}.  The corresponding
statistical mechanical model is known to be integrable, via
star-triangle relation, when these curves degenerate to rational
curves and the symmetry group becomes
finite~\cite{AuCoPeTaYa87,HaMa88,BaPeAu88,FaZa82}.

One should keep in mind that admissible patterns of size $5\times 5$
easily yield transformations with exponential growth of the
degree. For example the cyclic matrix (chiral Potts model, see
eq.~(\ref{genepotts})) yields a map with non vanishing entropy
\begin{eqnarray}
\label{gold2}
\epsilon = \log ( { {7+3 \,\sqrt{5}}\over{2}}  ),
\end{eqnarray}
and the corresponding statistical mechanical model is generically not
star-triangle integrable.  

We have used the arithmetical method described in
section~(\ref{complexity}) to evaluate the entropy for both the
symmetric pattern (symmetric $5\times 5 $ Boltzmann matrix) and the
general five-state model ($5 \times 5$ Boltzmann matrix with no
relations between the entries)\footnote{For a generic $5 \times 5$
matrix the coefficients got as large as $10^{720000}$ after six
iterations.}  . We found {\em the same value} for the entropy as for
the cyclic matrix, given by equation~(\ref{gold2}).

In contrast, if one restricts the $5\times 5$ Boltzmann matrix to be
both cyclic and symmetric, the entropy vanishes, and the
transformation $\varphi$ has one algebraic
invariant~\cite{BeMaRoVi92}.

One may wonder if the entropy calculated for the cyclic matrix, the
symmetric matrix, and the most general $q\times q$ matrix still
identify for $q \geq 6$. One may also try to evaluate the entropy for
the cyclic-symmetric pattern, and provide a closed expression valid
for all values of $q$, and similar to~(\ref{gold2}).

\section{Conclusion and perspectives}

We have given a classification of four-state spin models according to
the properties of the realization of {\is}.

This preparatory work provides the optimal parametrizations of the
models.

If one is looking for integrability via solutions of star-triangle
equations, one should stick to cases where the parameters live on
invariant algebraic varieties. Those may be present for generic values
of the parameters, or appear by a collapse of the realization of the
infinite group to a finite group, or by the degeneration of
transcendental invariants to algebraic ones.

The algebraic approach presented here may be complemented with random
matrix theory, as in~\cite{AnMaVi02}, to locate solvability of the
lattice models.

One may have to examine in detail the properties of both the matrix
product and the element by element product, and not just the stability
by the corresponding inverses, when mixed with matrix transposition.
One should notice that the Boltzmann matrices of the models which are
known to be integrable via the star-triangle relation, i.e $\Gamma_1$,
$\Gamma_2$, $\Gamma_4$, $\Gamma_5$ with the notations of
section~\ref{extended}, are (bi)-stochastic matrices (sum of elements
on lines and columns are constant).

A detailed analysis of the star-triangle equations will then proceed
on a case by case basis, which is beyond the scope of this paper.


\appendix

\section{List of the 42 admissible patterns for $q=4$}
\label{appa}
The sixteen entries of each $4\times 4$ matrix are presented on one
line in the order $ [W_{11}, W_{12}, W_{13}, W_{14}, W_{21}, W_{22},
\dots,W_{43}, W_{44}]$. The choice of the representative is arbitrary,
and should be understood modulo permutations of lines and columns. The
list is arranged with increasing number of independent homogeneous
parameters, and this number is indicated after the list of entries.
The last item of each line is the number of patterns in the 
similarity class.
\begin{eqnarray*} \hskip -2truecm
{\#                                       1}:&
[{\it w_0},{\it w_0},{\it w_0},{\it w_1},{\it w_0},{\it w_0},{\it w_1},{\it 
w_0},{\it w_0},{\it w_1},{\it w_0},{\it w_0},{\it w_1},{\it w_0},{\it w_0},{
\it w_0}] &\quad 2 \;, \quad 3
\\ \hskip -2truecm
{\#                                       2}:&
[{\it w_0},{\it w_0},{\it w_1},{\it w_0},{\it w_0},{\it w_1},{\it w_0},{\it 
w_0},{\it w_1},{\it w_0},{\it w_0},{\it w_0},{\it w_0},{\it w_0},{\it w_0},{
\it w_1}] &\quad 2 \;, \quad 6
\\ \hskip -2truecm
{\#                                       3}:&
[{\it w_0},{\it w_1},{\it w_1},{\it w_1},{\it w_1},{\it w_0},{\it w_1},{\it 
w_1},{\it w_1},{\it w_1},{\it w_0},{\it w_1},{\it w_1},{\it w_1},{\it w_1},{
\it w_0}] &\quad 2 \;, \quad 1
\\ \hskip -2truecm
{\#                                       4}:&
[{\it w_0},{\it w_1},{\it w_2},{\it w_2},{\it w_1},{\it w_0},{\it w_2},{\it 
w_2},{\it w_2},{\it w_2},{\it w_1},{\it w_0},{\it w_2},{\it w_2},{\it w_0},{
\it w_1}] &\quad 3 \;, \quad 3
\\ \hskip -2truecm
{\#                                       5}:&
[{\it w_0},{\it w_1},{\it w_2},{\it w_0},{\it w_1},{\it w_0},{\it w_0},{\it 
w_2},{\it w_2},{\it w_0},{\it w_0},{\it w_1},{\it w_0},{\it w_2},{\it w_1},{
\it w_0}] &\quad 3 \;, \quad 3
\\ \hskip -2truecm
{\#                                       6}:&
[{\it w_0},{\it w_0},{\it w_1},{\it w_2},{\it w_0},{\it w_0},{\it w_2},{\it 
w_1},{\it w_2},{\it w_1},{\it w_0},{\it w_0},{\it w_1},{\it w_2},{\it w_0},{
\it w_0}] &\quad 3 \;, \quad 3
\\ \hskip -2truecm
{\#                                       7}:&
[{\it w_0},{\it w_1},{\it w_2},{\it w_2},{\it w_1},{\it w_0},{\it w_2},{\it 
w_2},{\it w_2},{\it w_2},{\it w_0},{\it w_1},{\it w_2},{\it w_2},{\it w_1},{
\it w_0}] &\quad 3 \;, \quad 3
\\ \hskip -2truecm
{\#                                       8}:&
[{\it w_0},{\it w_1},{\it w_1},{\it w_1},{\it w_1},{\it w_2},{\it w_3},{\it 
w_3},{\it w_1},{\it w_3},{\it w_2},{\it w_3},{\it w_1},{\it w_3},{\it w_3},{
\it w_2}] &\quad 4 \;, \quad 4
\\ \hskip -2truecm
{\#                                       9}:&
[{\it w_0},{\it w_1},{\it w_1},{\it w_1},{\it w_1},{\it w_2},{\it w_3},{\it 
w_2},{\it w_1},{\it w_3},{\it w_2},{\it w_2},{\it w_1},{\it w_2},{\it w_2},{
\it w_3}] &\quad 4 \;, \quad 12
\\ \hskip -2truecm
{\#                                       10}:&
[{\it w_0},{\it w_1},{\it w_2},{\it w_3},{\it w_1},{\it w_0},{\it w_3},{\it 
w_2},{\it w_2},{\it w_3},{\it w_1},{\it w_0},{\it w_3},{\it w_2},{\it w_0},{
\it w_1}] &\quad 4 \;, \quad 3
\\ \hskip -2truecm
{\#                                       11}:&
[{\it w_0},{\it w_1},{\it w_2},{\it w_3},{\it w_1},{\it w_0},{\it w_3},{\it 
w_2},{\it w_3},{\it w_2},{\it w_1},{\it w_0},{\it w_2},{\it w_3},{\it w_0},{
\it w_1}] &\quad 4 \;, \quad 3
\\ \hskip -2truecm
{\#                                       12}:&
[{\it w_0},{\it w_1},{\it w_1},{\it w_2},{\it w_3},{\it w_2},{\it w_0},{\it 
w_3},{\it w_3},{\it w_0},{\it w_2},{\it w_3},{\it w_2},{\it w_1},{\it w_1},{
\it w_0}] &\quad 4 \;, \quad 3
\\ \hskip -2truecm
{\#                                       13}:&
[{\it w_0},{\it w_1},{\it w_0},{\it w_2},{\it w_2},{\it w_3},{\it w_1},{\it 
w_3},{\it w_0},{\it w_2},{\it w_0},{\it w_1},{\it w_1},{\it w_3},{\it w_2},{
\it w_3}] &\quad 4 \;, \quad 3
\\ \hskip -2truecm
{\#                                       14}:&
[{\it w_0},{\it w_1},{\it w_2},{\it w_0},{\it w_1},{\it w_3},{\it w_3},{\it 
w_2},{\it w_2},{\it w_3},{\it w_3},{\it w_1},{\it w_0},{\it w_2},{\it w_1},{
\it w_0}] &\quad 4 \;, \quad 3
\\ \hskip -2truecm
{\#                                       15}:&
[{\it w_0},{\it w_1},{\it w_1},{\it w_2},{\it w_3},{\it w_0},{\it w_2},{\it 
w_3},{\it w_3},{\it w_2},{\it w_0},{\it w_3},{\it w_2},{\it w_1},{\it w_1},{
\it w_0}] &\quad 4 \;, \quad 3
\\ \hskip -2truecm
{\#                                       16}:&
[{\it w_0},{\it w_1},{\it w_2},{\it w_3},{\it w_1},{\it w_0},{\it w_3},{\it 
w_2},{\it w_2},{\it w_3},{\it w_0},{\it w_1},{\it w_3},{\it w_2},{\it w_1},{
\it w_0}] &\quad 4 \;, \quad 1
\\ \hskip -2truecm
{\#                                       17}:&
[{\it w_0},{\it w_1},{\it w_2},{\it w_3},{\it w_1},{\it w_0},{\it w_3},{\it 
w_2},{\it w_3},{\it w_2},{\it w_0},{\it w_1},{\it w_2},{\it w_3},{\it w_1},{
\it w_0}] &\quad 4 \;, \quad 3
\\ \hskip -2truecm
{\#                                       18}:&
[{\it w_0},{\it w_1},{\it w_2},{\it w_0},{\it w_3},{\it w_4},{\it w_3},{\it 
w_3},{\it w_2},{\it w_1},{\it w_0},{\it w_0},{\it w_0},{\it w_1},{\it w_0},{
\it w_2}] &\quad 5 \;, \quad 12
\\ \hskip -2truecm
{\#                                       19}:&
[{\it w_0},{\it w_1},{\it w_2},{\it w_3},{\it w_1},{\it w_4},{\it w_1},{\it 
w_1},{\it w_3},{\it w_1},{\it w_0},{\it w_2},{\it w_2},{\it w_1},{\it w_3},{
\it w_0}] &\quad 5 \;, \quad 4
\\ \hskip -2truecm
{\#                                       20}:&
[{\it w_0},{\it w_1},{\it w_2},{\it w_0},{\it w_3},{\it w_0},{\it w_0},{\it 
w_4},{\it w_4},{\it w_0},{\it w_0},{\it w_3},{\it w_0},{\it w_2},{\it w_1},{
\it w_0}] &\quad 5 \;, \quad 3
\\ \hskip -2truecm
{\#                                       21}:&
[{\it w_0},{\it w_1},{\it w_2},{\it w_1},{\it w_3},{\it w_2},{\it w_4},{\it 
w_0},{\it w_2},{\it w_1},{\it w_0},{\it w_1},{\it w_4},{\it w_0},{\it w_3},{
\it w_2}] &\quad 5 \;, \quad 6
\\ \hskip -2truecm
{\#                                       22}:&
[{\it w_0},{\it w_1},{\it w_1},{\it w_1},{\it w_2},{\it w_3},{\it w_4},{\it 
w_4},{\it w_2},{\it w_4},{\it w_3},{\it w_4},{\it w_2},{\it w_4},{\it w_4},{
\it w_3}] &\quad 5 \;, \quad 4
\\ \hskip -2truecm
{\#                                       23}:&
[{\it w_0},{\it w_1},{\it w_2},{\it w_2},{\it w_1},{\it w_0},{\it w_2},{\it 
w_2},{\it w_2},{\it w_2},{\it w_3},{\it w_4},{\it w_2},{\it w_2},{\it w_4},{
\it w_3}] &\quad 5 \;, \quad 3
\\ \hskip -2truecm
{\#                                       24}:&
[{\it w_0},{\it w_1},{\it w_1},{\it w_1},{\it w_1},{\it w_2},{\it w_3},{\it 
w_4},{\it w_1},{\it w_3},{\it w_4},{\it w_2},{\it w_1},{\it w_4},{\it w_2},{
\it w_3}] &\quad 5 \;, \quad 4
\\ \hskip -2truecm
{\#                                       25}:&
[{\it w_0},{\it w_1},{\it w_2},{\it w_3},{\it w_4},{\it w_0},{\it w_4},{\it 
w_2},{\it w_2},{\it w_3},{\it w_0},{\it w_1},{\it w_4},{\it w_2},{\it w_4},{
\it w_0}] &\quad 5 \;, \quad 6
\\ \hskip -2truecm
{\#                                       26}:&
[{\it w_0},{\it w_1},{\it w_2},{\it w_3},{\it w_4},{\it w_0},{\it w_3},{\it 
w_5},{\it w_5},{\it w_3},{\it w_0},{\it w_4},{\it w_3},{\it w_2},{\it w_1},{
\it w_0}] &\quad 6 \;, \quad 3
\\ \hskip -2truecm
{\#                                       27}:&
[{\it w_0},{\it w_1},{\it w_0},{\it w_2},{\it w_3},{\it w_4},{\it w_5},{\it 
w_4},{\it w_0},{\it w_2},{\it w_0},{\it w_1},{\it w_5},{\it w_4},{\it w_3},{
\it w_4}] &\quad 6 \;, \quad 3
\\ \hskip -2truecm
{\#                                       28}:&
[{\it w_0},{\it w_1},{\it w_1},{\it w_1},{\it w_2},{\it w_3},{\it w_4},{\it 
w_5},{\it w_2},{\it w_5},{\it w_3},{\it w_4},{\it w_2},{\it w_4},{\it w_5},{
\it w_3}] &\quad 6 \;, \quad 4
\\ \hskip -2truecm
{\#                                       29}:&
[{\it w_0},{\it w_1},{\it w_2},{\it w_1},{\it w_3},{\it w_4},{\it w_3},{\it 
w_5},{\it w_2},{\it w_1},{\it w_0},{\it w_1},{\it w_3},{\it w_5},{\it w_3},{
\it w_4}] &\quad 6 \;, \quad 3
\\ \hskip -2truecm
{\#                                       30}:&
[{\it w_0},{\it w_1},{\it w_2},{\it w_3},{\it w_1},{\it w_4},{\it w_3},{\it 
w_5},{\it w_2},{\it w_3},{\it w_0},{\it w_1},{\it w_3},{\it w_5},{\it w_1},{
\it w_4}] &\quad 6 \;, \quad 3
\\ \hskip -2truecm
{\#                                       31}:&
[{\it w_0},{\it w_1},{\it w_2},{\it w_3},{\it w_3},{\it w_4},{\it w_1},{\it 
w_5},{\it w_2},{\it w_3},{\it w_0},{\it w_1},{\it w_1},{\it w_5},{\it w_3},{
\it w_4}] &\quad 6 \;, \quad 3
\\ \hskip -2truecm
{\#                                       32}:&
[{\it w_0},{\it w_1},{\it w_1},{\it w_1},{\it w_2},{\it w_3},{\it w_4},{\it 
w_5},{\it w_2},{\it w_4},{\it w_5},{\it w_3},{\it w_2},{\it w_5},{\it w_3},{
\it w_4}] &\quad 6 \;, \quad 4
\\ \hskip -2truecm
{\#                                       33}:&
[{\it w_0},{\it w_1},{\it w_2},{\it w_3},{\it w_4},{\it w_2},{\it w_5},{\it 
w_0},{\it w_2},{\it w_3},{\it w_0},{\it w_1},{\it w_5},{\it w_0},{\it w_4},{
\it w_2}] &\quad 6 \;, \quad 3
\\ \hskip -2truecm
{\#                                       34}:&
[{\it w_0},{\it w_1},{\it w_2},{\it w_1},{\it w_1},{\it w_3},{\it w_4},{\it 
w_5},{\it w_2},{\it w_4},{\it w_6},{\it w_4},{\it w_1},{\it w_5},{\it w_4},{
\it w_3}] &\quad 7 \;, \quad 6
\\ \hskip -2truecm
{\#                                       35}:&
[{\it w_0},{\it w_1},{\it w_2},{\it w_2},{\it w_1},{\it w_0},{\it w_2},{\it 
w_2},{\it w_3},{\it w_4},{\it w_5},{\it w_6},{\it w_4},{\it w_3},{\it w_6},{
\it w_5}] &\quad 7 \;, \quad 6
\\ \hskip -2truecm
{\#                                       36}:&
[{\it w_0},{\it w_1},{\it w_2},{\it w_2},{\it w_3},{\it w_0},{\it w_4},{\it 
w_4},{\it w_4},{\it w_2},{\it w_5},{\it w_6},{\it w_4},{\it w_2},{\it w_6},{
\it w_5}] &\quad 7 \;, \quad 6
\\ \hskip -2truecm
{\#                                       37}:&
[{\it w_0},{\it w_1},{\it w_2},{\it w_3},{\it w_1},{\it w_0},{\it w_3},{\it 
w_2},{\it w_4},{\it w_5},{\it w_6},{\it w_7},{\it w_5},{\it w_4},{\it w_7},{
\it w_6}] &\quad 8 \;, \quad 3
\\ \hskip -2truecm
{\#                                       38}:&
[{\it w_0},{\it w_1},{\it w_2},{\it w_2},{\it w_3},{\it w_4},{\it w_5},{\it 
w_5},{\it w_6},{\it w_7},{\it w_8},{\it w_9},{\it w_6},{\it w_7},{\it w_9},{
\it w_8}] &\quad 10 \;, \quad 6
\\ \hskip -2truecm
{\#                                       39}:&
[{\it w_0},{\it w_1},{\it w_2},{\it w_3},{\it w_1},{\it w_4},{\it w_5},{\it 
w_6},{\it w_2},{\it w_5},{\it w_7},{\it w_8},{\it w_3},{\it w_6},{\it w_8},{
\it w_9}] &\quad 10 \;, \quad 1
\\ \hskip -2truecm
{\#                                       40}:&
[{\it w_0},{\it w_1},{\it w_2},{\it w_3},{\it w_4},{\it w_0},{\it w_5},{\it 
w_6},{\it w_6},{\it w_3},{\it w_7},{\it w_8},{\it w_5},{\it w_2},{\it w_9},{
\it w_7}] &\quad 10 \;, \quad 3
\\ \hskip -2truecm
{\#                                       41}:&
[{\it w_0},{\it w_1},{\it w_2},{\it w_3},{\it w_4},{\it w_5},{\it w_1},{\it 
w_6},{\it w_7},{\it w_4},{\it w_0},{\it w_8},{\it w_8},{\it w_6},{\it w_3},{
\it w_9}] &\quad 10 \;, \quad 6
\\ \hskip -2truecm
{\#                                       42}:&
[{\it w_0},{\it w_1},{\it w_2},{\it w_3},{\it w_4},{\it w_5},{\it w_6},{\it 
w_7},{\it w_8},{\it w_9},{\it w_{10}},{\it w_{11}},{\it w_{12}},{\it w_{13}},{\it w_{14}
},{\it w_{15}}] &\quad 16 \;, \quad 1
\end{eqnarray*}

\section{Generating functions}
\label{generating}
\begin{eqnarray*}
&\mbox{ Pattern number } \qquad & \mbox{ Generating function }  \\
& 1, 2, 3 & {}{1\over{1-s}}{} \\
& 4, 5, 6, 7 & {}{\frac { \left( 1+s \right) ^{2}}{ \left( 1-s \right) ^{2}}}{}
\\
& 8, 9 & {}{\frac { \left( 1+s \right)  \left( 1+2\,s-{s}^{2}+{s}^{4} \right) }{ \left( 1-s \right) ^{2}}}{} \\
& 10, 11, 16, 17 & 
{}{\frac { \left( 1+3\,s \right) ^{2}}{ \left( 1-s \right) ^{3}}}{}\\
& 12, 13, 14, 15& 
{}{\frac {1+3\,s}{ \left( 1-s \right) ^{2}}}{} \\
& 18, 22& 
{}{\frac {1+4\,s-{s}^{2}+{s}^{4}+{s}^{5}}{ \left( 1-s \right) ^{2}}}{}\\
& 19, 24& 
{}{\frac { \left( 1+s \right)  \left( 1+6\,s+3\,{s}^{2}-{s}^{3} \right) 
}{ \left( 1-s \right) ^{3}}}{} \\
& 20, 23&
{}{\frac { \left( 1+s \right)  \left( 1+4\,s-{s}^{2} \right) }{ \left( 1
-s \right) ^{3}}}{}\\
& 21, 25& 
{}{\frac {1+8\,s+12\,{s}^{2}+18\,{s}^{3}+13\,{s}^{4}+6\,{s}^{5}-2\,{s}^{
6}}{ \left( 1+s \right)  \left( 1+s+{s}^{2} \right)  \left( 1-s
 \right) ^{4}}}{} \\
& 26, 30, 31, 33 &
{}{\frac {1+11\,s+11\,{s}^{2}-7\,{s}^{3}}{ \left( 1-s \right) ^{4}}}{} \\
& 27, 29& 
{} {\frac {1+5\,s+4\,{s}^{2}-3\,{s}^{3}+{s}^{4}}{ \left( 1-s \right) ^{3}
}}{}\\
& 28, 32&
{}{\frac { \left( 1+s+{s}^{2} \right)  \left( 1+7\,s-2\,{s}^{2} \right) 
}{ \left( 1-s \right) ^{3}}}{}\\
& 34, 36& 
{}{\frac {1+12\,s+16\,{s}^{2}+3\,{s}^{3}+8\,{s}^{4}+4\,{s}^{5}-2\,{s}^{6
}}{ \left( 1+s \right)  \left( 1-s \right) ^{3}}}{} \\
& 35 & 
{}{\frac {1+12\,s+24\,{s}^{2}+34\,{s}^{3}+27\,{s}^{4}+14\,{s}^{5}}{
 \left( 1+s \right)  \left( 1+s+{s}^{2} \right)  \left( 1-s \right) ^{
4}}}{} \\
& 37 & 
{}{\frac {1+17\,s+19\,{s}^{2}-5\,{s}^{3}}{ \left( 1-s \right) ^{4}}}{} \\
& 38 &
{}{\frac {1+18\,s+11\,{s}^{2}+10\,{s}^{3}+8\,{s}^{4}+8\,{s}^{5}}{
 \left( 1+s \right)  \left( 1-s \right) ^{3}}}{}\\
& 39, 40, 41 &
{}{\frac {1+23\,s+15\,{s}^{2}+5\,{s}^{3}+4\,{s}^{4}}{ \left( 1-s
 \right) ^{4}}} {}\\
& 42 & 
{}{\frac {1+41\,s+3\,{s}^{2}+35\,{s}^{3}+16\,{s}^{4}}{ \left( 1-s
 \right) ^{4}}}{}
\end{eqnarray*}

\section{Admissible patterns for $q=3$}
\label{three}
\begin{eqnarray*}
{\#                                       1}:&
[{\it w_0},{\it w_1},{\it w_1},{\it w_1},{\it w_0},{\it w_1},{\it w_1},{\it 
w_1},{\it w_0}]
 &\quad  \;  \quad 
\\
{\#                                       2}:&
[{\it w_0},{\it w_0},{\it w_1},{\it w_0},{\it w_1},{\it w_0},{\it w_1},{\it 
w_0},{\it w_0}]
 &\quad  \;  \quad 
\\
{\#                                       3}:&
[{\it w_0},{\it w_1},{\it w_2},{\it w_2},{\it w_0},{\it w_1},{\it w_1},{\it 
w_2},{\it w_0}]
 &\quad  \;  \quad 
\\
{\#                                       4}:&
[{\it w_0},{\it w_1},{\it w_2},{\it w_1},{\it w_2},{\it w_0},{\it w_2},{\it 
w_0},{\it w_1}]
 &\quad  \;  \quad 
\\
{\#                                       5}:&
[{\it w_0},{\it w_1},{\it w_1},{\it w_1},{\it w_2},{\it w_3},{\it w_1},{\it 
w_3},{\it w_2}]
 &\quad  \;  \quad 
\\
{\#                                       6}:&
[{\it w_0},{\it w_1},{\it w_1},{\it w_2},{\it w_3},{\it w_4},{\it w_2},{\it 
w_4},{\it w_3}]
 &\quad  \;  \quad 
\\
{\#                                       7}:&
[{\it w_0},{\it w_1},{\it w_2},{\it w_2},{\it w_3},{\it w_4},{\it w_1},{\it 
w_5},{\it w_3}]
 &\quad  \;  \quad 
\\
{\#                                       8}:&
[{\it w_0},{\it w_1},{\it w_2},{\it w_1},{\it w_3},{\it w_4},{\it w_2},{\it 
w_4},{\it w_5}]
 &\quad  \;  \quad 
\\
{\#                                       9}:&
[{\it w_0},{\it w_1},{\it w_2},{\it w_3},{\it w_4},{\it w_5},{\it w_6},{\it 
w_7},{\it w_8}]
 &\quad  \;  \quad 
\end{eqnarray*}

\section{Some algebraic invariants for $q=3$} 
\label{inva_three}
The lower index is the pattern number.
\begin{eqnarray*}
&\Delta_5^{(a)} =&
{\frac { \left({{w_1}}^{2}{w_3} -{w_0}\,{{w_2}}^{2}
 \right) ^{2}}{{w_2}\, \left({{w_1}}^{2} -{w_0}\,{w_2}
 \right)  \left( {w_2}\,{{w_1}}^{2}-{w_0}\,{{w_3}}^{2}
 \right) }},\\ \qquad
&\Delta_5^{(b)} =& 
{\frac { \left( {w_2}\,{{w_1}}^{2}-{w_0}\,{{w_3}}^{2}
 \right) ^{2}}{{w_3}\, \left({{w_1}}^{2} -{w_0}\,{w_3}
 \right)  \left({{w_1}}^{2}{w_3} -{w_0}\,{{w_2}}^{2}
 \right) }}.
\end{eqnarray*}
\begin{eqnarray*}
&\Delta_6^{(a)} =& {\frac {{w_1}\,{w_2}}{{{w_2}}^{2}+{{w_1}}^{2}}},
\\ & \Delta_6^{(b)} =& {\frac { \left({w_1}\,{w_2}\,{w_4}
-{w_0}\,{{w_3}}^{2} \right) ^{2}}{{w_3}\, \left({w_1}\,{w_2 }
-{w_0}\,{w_3} \right) \left( {w_1}\,{w_2}\,{w_3}-{{w_4}}^{2}{w_0}
\right) }} ,\\ \qquad &\Delta_6^{(c)} =& {\frac { \left(
{w_1}\,{w_2}\,{w_3}-{{w_4}}^{2}{w_0} \right) ^{2}}{{w_4}\,
\left({w_1}\,{w_2 } -{w_0}\,{w_4} \right) \left({w_1}\,{w_2}\,{w_4}
-{w_0}\,{{w_3}}^{2} \right) }}.
\end{eqnarray*}
\begin{eqnarray*}
\hskip -2truecm
\Delta_8^{(a)} =
{\frac {{w_0}\, \left( {w_3}\,{w_5}-{{w_4}}^{2} \right) }
{ \left( {w_4}\,{w_2}-{w_1}\,{w_5} \right) {w_1}}}
 ,\qquad
 \Delta_8^{(b)} =
{\frac {{w_0}\, \left( {w_3}\,{w_5}-{{w_4}}^{2} \right) }{
 \left( {w_1}\,{w_4}-{w_2}\,{w_3} \right) {w_2}}}
 ,\qquad
\Delta_8^{(c)} =
{\frac {{w_2}\,{w_3}}{{w_1}\,{w_4}}}.
\end{eqnarray*}

\bigskip


\end{document}